\documentclass[lettersize,journal]{IEEEtran}
  \usepackage[nocompress]{cite}
\ifCLASSINFOpdf
  \usepackage[pdftex]{graphicx}
\else
  \usepackage[dvips]{graphicx}
\fi
\usepackage{amsmath}
\usepackage{array}

\usepackage[caption=false,font=normalsize,labelfont=sf,textfont=sf]{subfig}
\usepackage{fixltx2e}

\usepackage{stfloats}
\usepackage{url}

\usepackage{amsmath,amsfonts,bm}

\def\eqref#1{equation~\ref{#1}}

\def\1{\bm{1}}

\def\rvx{{\mathbf{x}}}
\def\rvy{{\mathbf{y}}}
\def\rvz{{\mathbf{z}}}

\def\ervy{{\textnormal{y}}}

\def\rmE{{\mathbf{E}}}

\def\rmG{{\mathbf{G}}}

\def\rmL{{\mathbf{L}}}

\def\rmN{{\mathbf{N}}}

\def\rmV{{\mathbf{V}}}

\def\rmX{{\mathbf{X}}}

\def\ermT{{\textnormal{T}}}

\def\vtheta{{\bm{\theta}}}

\def\vx{{\bm{x}}}

\def\mG{{\bm{G}}}

\def\mT{{\bm{T}}}

\def\mPhi{{\bm{\Phi}}}

\DeclareMathAlphabet{\mathsfit}{\encodingdefault}{\sfdefault}{m}{sl}
\SetMathAlphabet{\mathsfit}{bold}{\encodingdefault}{\sfdefault}{bx}{n}
\newcommand{\tens}[1]{\bm{\mathsfit{#1}}}

\def\tB{{\tens{B}}}

\def\tW{{\tens{W}}}

\def\emT{{T}}

\newcommand{\etens}[1]{\mathsfit{#1}}

\def\etW{{\etens{W}}}

\newcommand{\E}{\mathbb{E}}
\newcommand{\Ls}{\mathcal{L}}
\newcommand{\R}{\mathbb{R}}

\DeclareMathOperator*{\argmax}{arg\,max}

\DeclareMathOperator{\categorical}{Categorical}

\DeclareMathOperator{\uniform}{Uniform}

\DeclareMathOperator{\dynconv}{dynconv}
\DeclareMathOperator{\concat}{Concat}

\DeclareMathOperator{\MLP}{MLP}
\DeclareMathOperator{\AR}{AR}
\DeclareMathOperator{\MAC}{MAC}

\usepackage{comment}
\usepackage{booktabs}
\usepackage{hyperref}
\usepackage{color}

\usepackage{multirow}
\usepackage{multicol}

\usepackage{tabularx}
\usepackage{algorithm}
\usepackage{algpseudocode}

\usepackage{balance}
\usepackage{comment}
\allowdisplaybreaks[4]

\newcommand{\zyf}[1]{\textcolor{black}{#1}}

\begin{document}
%
\title{ABC: Adaptive BayesNet Structure Learning for Computational Scalable Multi-task Image Compression}
%
%
%
%

\author{Yufeng~Zhang, Wenrui~Dai, Hang~Yu, Shizhan~Liu, Junhui~Hou, Jianguo~Li and Weiyao~Lin

\thanks{
Yufeng~Zhang, Wenrui~Dai, Shizhan~Liu and Weiyao~Lin were with the Department
of Electrical Engineering, Shanghai Jiao Tong University, Shanghai, China.
E-mail: \{worldlife,shanluzuode,daiwenrui,wylin\}@sjtu.edu.cn}

\thanks{Hang~Yu and Jianguo~Li were with the Ant Group, Hangzhou, China. 
E-mail: \{hyu.hugo,lijg.zero\}@antgroup.com}

\thanks{Junhui~Hou was with the Department of Computer Science, City University of Hong Kong, Kowloon, Hong Kong.
E-mail: jh.hou@cityu.edu.hk}

\thanks{Weiyao Lin is the Corresponding Author.
}

}

%
%

\markboth{IEEE TRANSACTIONS ON PATTERN ANALYSIS AND MACHINE INTELLIGENCE}%
{Zhang \MakeLowercase{\textit{et al.}}: ABC: Adaptive BayesNet Structure Learning for Computational Scalable Multi-task Image Compression}
%



\IEEEtitleabstractindextext{%
\begin{abstract}
Neural Image Compression (NIC) has revolutionized image compression with its superior rate-distortion performance and multi-task capabilities, supporting both human visual perception and machine vision tasks. However, its widespread adoption is hindered by substantial computational demands. While existing approaches attempt to address this challenge through module-specific optimizations or pre-defined complexity levels, they lack comprehensive control over computational complexity. We present ABC (Adaptive BayesNet structure learning for computational scalable multi-task image Compression), a novel, comprehensive framework that achieves computational scalability across all NIC components through Bayesian network (BayesNet) structure learning. ABC introduces three key innovations: (i) a heterogeneous bipartite BayesNet (inter-node structure) for managing neural backbone computations; (ii) a homogeneous multipartite BayesNet (intra-node structure) for optimizing autoregressive unit processing; and (iii) an adaptive control module that dynamically adjusts the BayesNet structure based on device capabilities, input data complexity, and downstream task requirements. Experiments demonstrate that ABC enables full computational scalability with better complexity adaptivity and broader complexity control span, while maintaining competitive compression performance. Furthermore, the framework's versatility allows integration with various NIC architectures that employ BayesNet representations, making it a robust solution for ensuring computational scalability in NIC applications. Code is available in \url{https://github.com/worldlife123/cbench_BaSIC}.

\end{abstract}

}

\maketitle

\IEEEdisplaynontitleabstractindextext

%
\IEEEpeerreviewmaketitle

\section{Introduction}
\label{sec:intro}

\begin{figure}
    \centering

    \subfloat[Typical Multi-task Image Compression Framework.]{
        \centering
        \includegraphics[width=\linewidth]{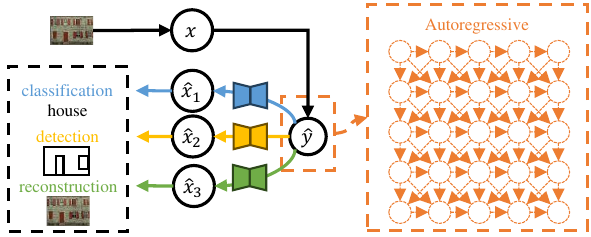}
        \label{fig:showcase-framework}
    }

    \subfloat[Adaptive Frameworks for Computational Scalability.]{
        \centering
        \includegraphics[width=\linewidth]{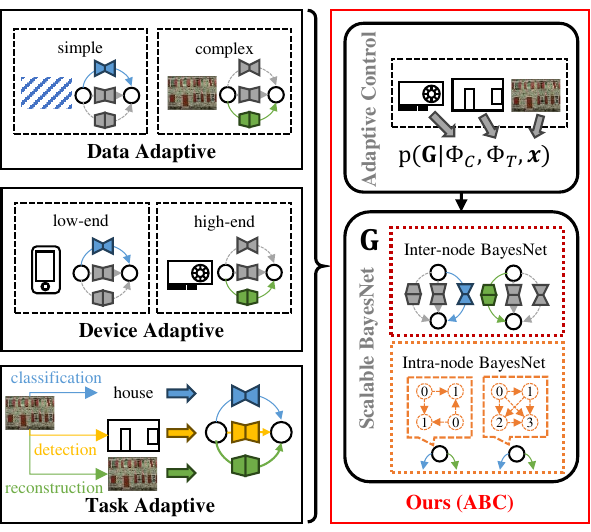}
        \label{fig:showcase-adaptive}
    }

    \caption{Computational Scalable Multi-task Image Compression.}
    \label{fig:showcase}
\end{figure}

Neural Image Compression (NIC)~\cite{Ball2018VariationalIC, Cheng2020LearnedIC, He2022ELICEL,Chen2023TransTICTT} represents a significant advancement in image compression, ushering in a transformative phase compared to traditional codecs like JPEG and WebP. Unlike its predecessors, NIC leverages deep neural networks to achieve superior rate-distortion performance while inherently supporting multi-task capabilities. These capabilities extend beyond human visual perception to encompass diverse machine vision tasks, such as image classification and instance segmentation\cite{Ascenso2023TheJA, Chen2023TransTICTT, Cao2023SlimmableMI, Li2024ImageCF}. 
A typical multi-task NIC framework, often depicted as a Bayesian network (BayesNet) in \autoref{fig:showcase-framework}, integrates autoregressive models for entropy coding and task-specific neural decoders, enabling adaptable decompression tailored to varied vision tasks.


Although NIC has immense potential, its practical deployment is still hindered by the substantial computational demands of the underlying NIC modules. Therefore, \textbf{Computational Scalability}, the ability to adjust computational complexity within the codec, is crucial.
Traditional image codecs typically accommodate this functionality by manually setting a complexity level. Indeed, WebP offers seven levels of computational complexity to regulate lossy image compression\cite{libwebp}.
However, such rigid control lacks the flexibility required for modern applications, where computational demands must adapt dynamically to factors such as device capabilities, input data complexity, and the requirements of downstream tasks. For instance, high-end devices may prioritize compression quality, while resource-constrained edge devices necessitate energy-efficient processing. Similarly, simple images with uniform textures demand less computational effort than complex scenes, and machine vision tasks often tolerate higher compression losses compared to human-centric applications. 
Recent advancements \cite{Yang2021SlimmableCA} have introduced dynamic neural networks, such as slimmable networks \cite{Yu2018SlimmableNN}, which offer a more efficient approach. These networks can dynamically adjust their computational complexity within a single model, eliminating the need for multiple independent networks.
Additionally, to address data variability, input-adaptive complexity allocation modules for slimmable networks have been developed~\cite{Tao2023AdaNICTP, Zhang2023ELFICAL}. Finally,
regarding different vision tasks, Cao et al. \cite{Cao2023SlimmableMI} proposed controlling slimmable networks within the encoder to produce distinct bitstreams tailored to different tasks.
Nevertheless, these existing approaches have only partially addressed computational scalability with adaptive control based on limited conditions. Moreover, they often neglect crucial components in NIC, such as the essential autoregressive modules, which limits their computation control span.

In this work, we endeavor to create a novel framework that unifies computational scalability with adaptive control across all NIC components, including both the neural backbone and autoregressive units, as illustrated in \autoref{fig:showcase-framework}. 
We term this framework \textbf{ABC} (\textbf{A}daptive \textbf{B}ayesNet structure learning for computational scalable multi-task image \textbf{C}ompression).
Our advancement is a NIC framework that exploits end-to-end BayesNet structure learning\cite{Yu2019DAGGNNDS, Lorch2021DiBSDB} to meticulously optimize computational scalability. Crucially, this innovative framework is not tied to specific neural network backbones, making it versatile enough to enhance computational scalability in most existing NIC frameworks with a hyperprior BayesNet representation. 

Specifically, we represent the BayesNet structure as $\rmG$, where the nodes correspond to the input and latent information, and the edges represent conditional dependencies between nodes. Given that the structure we seek to learn must conform to a directed acyclic graph (DAG), we tailor the prior for compression frameworks within this DAG constraint, considering two distinct instances: the \textbf{inter-node BayesNet}, representing neural networks to capture conditional dependencies between nodes, and the \textbf{intra-node BayesNet}, which pertains to autoregressive model that models dependencies within nodes. 
Moreover, we introduce an \textbf{Adaptive Control Module} to tailor the BayesNet structure based on the following three factors:
\begin{itemize}
    \item \textbf{Device Adaptivity} adjusts codec performance to specific computational and memory constraints of different devices. High-end devices can afford more complex computations to achieve superior encoding quality, whereas low-end devices prioritize minimizing computational load for energy and resource efficiency.
    \item \textbf{Data Adaptivity} allocates computational effort based on the complexity of input data. Specifically, simple images, such as those with uniform colors or repetitive patterns, can be effectively compressed using lightweight codecs. Conversely, complex images require more intensive processing to preserve  details. 
    \item \textbf{Task Adaptivity} optimizes compression for different downstream tasks. For example, images intended for human vision typically require preservation of fine-grained details, while simpler machine vision tasks may tolerate greater loss of detail.
\end{itemize}

Mathematically, we impose a conditional prior $p(\rmG | \Phi_{C}, \Phi_{T}, \vx)$ on $\rmG$, where $\Phi_{C}$ is a controller node determined by the computation budget, $\Phi_{T}$ is a controller node determined by the vision task, and $\vx$ is the input data.

To summarize, our overall contributions are concluded as:
\begin{itemize}
\item We propose an end-to-end computationally scalable NIC framework based on BayesNet structure learning, which allows for fine-tuned and adaptive adjustments to computational complexity across all facets of the NIC process.
\item We propose \textbf{Heterogeneous bipartite structure for inter-node BayesNet}, which harnesses existing dynamic neural networks for effective computational scalability.
\item We propose \textbf{Homogeneous multipartite graph for intra-node BayesNet}, tailored for parallel computation in practical applications and enhancing compression performance.
\item We design \textbf{Adaptive Control Module} to generate the BayesNet structure, enabling computational scalability with device adaptivity, data adaptivity and vision task adaptivity.
\end{itemize}


This work builds upon our prior conference paper~\cite{Zhang2024BaSIC}. Expanding on the conference version, our key additional contribution is the integration of multiple adaptive control sources into the computationally scalable image compression framework. Specifically, we now account for device computational capability, data content, and vision task requirements, enabling multi-task image compression. To this end, we have reformulated the BayesNet structure learning problem with an emphasis on adaptive controllers. 
Our proposed Adaptive Control Module dynamically adjusts the BayesNet structure to cater to various adaptivity requirements. We have included additional experiments that demonstrate the effectiveness of the Adaptive Control Module. These experiments showcase image compression under varying computational budgets (device adaptive control), instance-level computational scalability (data adaptive control), and computational scalability tailored for different machine vision tasks (task adaptive control).
Furthermore, we have developed a two-stage stable optimization strategy to mitigate the large search space of the BayesNet structure learning problem and further improve the compression performance.
Finally, to provide further validation, we have performed extra experiments, including visualization of the reconstructed images under different computational complexity levels or subsequent vision tasks. We also include ablation studies using various network implementations to demonstrate the robustness of our design choices.

The rest of this paper is organized as follows. \autoref{sec:rw} introduces existing researches regarding NIC and BayesNet structure learning. \autoref{sec:overview} begins to interpret NIC from the BayesNet view, introduce the proposed ABC framework, defining the optimization model and proposing the corresponding optimizer. \autoref{sec:method} delves into two critical challenges outlined in \autoref{sec:overview}, namely the structure learning problem of inter-node and intra-node BayesNet. \autoref{sec:impl} introduces an implementation of the proposed ABC framework in \autoref{sec:overview} based on neural networks adopted by existing NIC frameworks. \autoref{sec:exp} includes experimental results to show the effectiveness of the proposed framework. Finally, \autoref{sec:conc} offers concluding remarks.

\section{Related Work}
\label{sec:rw}

In this section, we briefly review recent works regarding NIC, its computational complexity problem, as well as BayesNet structure learning.

\subsection{Neural Image Compression}
\label{sec:rw-nic}

Neural image compression, as the name suggests, are image compression methods based on neural networks. 
In NIC frameworks, neural networks are typically trained in an end-to-end fashion using a rate-distortion loss function \cite{Ball2016EndtoendOI}, which optimizes the trade-off between the size of the compressed bitstream and the quality of the reconstructed image.
Subsequent advancements have focused on refining compression by using better generative models for entropy coding, such as hyperprior VAE\cite{Ball2018VariationalIC} and autoregressive models\cite{Minnen2018JointAA, Lee2018ContextadaptiveEM, Cheng2020LearnedIC}.
Other works designs better neural networks to improve the overall rate-distortion performance, including generalized divisive normalization(GDN)\cite{Ball2015DensityMO}, non-local attention modules\cite{Chen2019EndtoEndLI}, transformers\cite{Lu2021TransformerbasedIC}, etc.
Since the proposal of Video Coding for Machine\cite{Duan2020VideoCF},  there has been a growing trend in Neural Image Compression (NIC) research towards developing codecs that cater to multiple machine vision tasks, such as image classification, object detection, and instance segmentation.
Several studies\cite{Codevilla2021LearnedIC, Chamain2021EndtoendOI} have explored the design of specialized decoders, which could be trained to adapt to different machine vision tasks during decompression.
Alternatively, some other works\cite{Liu2022ImprovingMM, Chen2023TransTICTT, Li2024ImageCF} have introduced modules aimed at adjusting both the encoder and decoder to better suit the requirements of different tasks. 
While the aforementioned improvements offer better image compression, they often overlook the increased computational demands. To apply NIC in practice, some recent works are now addressing the need to accelerate these models.
First, to circumvent the slow serial iteration steps of masked convolutions on GPUs, newer studies introduce parallelized autoregressive models for GPU acceleration. For instance, \cite{Minnen2020ChannelWiseAE} presents a channel-wise parallel autoregressive model, while \cite{He2021CheckerboardCM} proposes a spatially parallel model using a checkerboard mask, requiring just two parallel stages. This concept is extended to a multistage model by \cite{Lin2023MultistageSC} for improvement on compression. Further, \cite{He2022ELICEL} amalgamates channel-wise and spatially parallel autoregressive models for better compression performance. Alternatively, \cite{Kang2022PILCPI, Zhang2024FiniteStateAE} design simplified autoregressive models, enabling high efficiency even without GPUs.
Yet, such models are typically hand-crafted and may not represent the most efficient trade-off between compression complexity and performance.
Alternatively, considering that neural networks (i.e., edges in the BayesNets) represent a significant portion of NIC's computational complexity, some works focus on accelerating these directly. For example, there are designs for more efficient neural network operators like a simplified GDN\cite{Johnston2019ComputationallyEN} and CPU-friendly down/up-samplers\cite{Zheng2021GetTB}. Others apply general neural network acceleration techniques within compression models, such as MorphNet\cite{Johnston2019ComputationallyEN} and learnable channel pruning\cite{Yin2021UniversalEV, Yin2022ExploringSS}.



Apart from acceleration, a handful of recent NIC works have also noticed the importance of computational scalability for different computation devices. 
Some works\cite{Johnston2019ComputationallyEN, Yin2022ExploringSS} implement variable complexity by optimizing the complexity together with the rate-distortion loss function, and computational scalability could be achieved by training models with different weight on the complexity regularizer. 
Recently, SlimCAE\cite{Yang2021SlimmableCA} employs slimmable networks\cite{Yu2018SlimmableNN} as the framework's backbone, enabling dynamic adjustments of bit-rate and computation through channel width modulation. Later developments\cite{Zhang2023ELFICAL} adopt universal slimmable networks for more precise scalability. 
Nonetheless, these approaches focus solely on the scalability of neural networks, largely disregarding the autoregressive models' scalability.


Recently, some works began to explore the possibility for data-adaptive computation control. For example, \cite{Tao2023AdaNICTP, Zhang2023ELFICAL} adopt slimmable models for adjusting computational complexity with the help of extra input-adaptive complexity allocation modules. Despite these advancements, the scope of computational scalability remains largely confined to neural network architectures.
Note that the idea of data-adaptive compression has already been implemented in many NIC frameworks with totally different intuitions, For instance, techniques like inference-time overfitting \cite{Campos2019ContentAO} and content-adaptive masks applied to the latent space \cite{Pan2022ContentAL, Zhang2023ContentAC} are primarily geared towards improving the fidelity and effectiveness of compressed data, rather than reducing the computational burden associated with the compression process.

Moreover, recent research has also ventured into task-adaptive computation control. Similar to data-adaptive frameworks, Cao et al. \cite{Cao2023SlimmableMI} utilize slimmable models with variable channel widths to produce distinct bitstreams tailored for different vision tasks. However, the adaptability of these models is constrained in two key aspects: channel widths can only be adjusted based on specific tasks rather than computational budget considerations, and modifications are limited to the encoder's channel widths alone.

\subsection{BayesNet Structure Learning}
\label{sec:rw-bnsl}

BayesNet structure learning aims to discern the conditional relationships among a set of variables with a graph representation, which ideally should be a directed acyclic graph (DAG). Traditional approaches often involve score-based\cite{Chickering2003OptimalSI, Yuan2013LearningOB} and constraint-based methods\cite{Cheng2002LearningBN, Koivisto2004ExactBS}. However, they typically operate within a constrained search space, which can lead to less-than-optimal graphs. More recent initiatives introduce neural network techniques for more robust graph generation within the realm of BayesNet structure learning. Some align this learning task with variational inference to facilitate graph structure development within an end-to-end framework\cite{Yu2019DAGGNNDS, Lorch2021DiBSDB}, utilizing graph acyclicity regularizers\cite{Zheng2018DAGsWN} that may not guarantee acyclicity. Alternatively, other research employs recurrent-style graph generators that assure acyclicity through iterations\cite{Deleu2022BayesianSL}, but these can become inefficient with an increase in the number of nodes.
In our work, we diverge from these established methods due to their inefficiency when dealing with the dense pixelation of images. Instead, we analyze the BayesNet representations inherent in existing NIC frameworks and directly parameterize the graph structure as DAGs, ensuring both acyclicity and efficiency.

\section{Overview of ABC}
\label{sec:overview}

\subsection{Bayesian Paradigm for NIC frameworks}

\begin{figure*}[t]
    \centering
    \begin{minipage}[]{\textwidth}
         \subfloat[Hyperprior\cite{Ball2018VariationalIC}]{
         \centering
         \includegraphics[width=.15\textwidth]{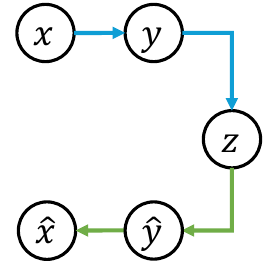}
         \label{fig:intro-hp}
         }
         \subfloat[SlimCAE\cite{Yang2021SlimmableCA}]{
             \centering
             \includegraphics[width=.15\textwidth]{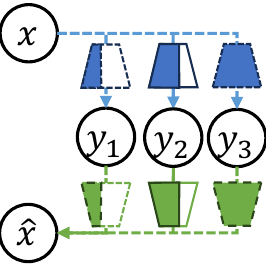}
             \label{fig:intro-slim}
         }
         \subfloat[Joint Autoregressive\cite{Minnen2018JointAA}]{
             \centering
             \includegraphics[width=.3\textwidth]{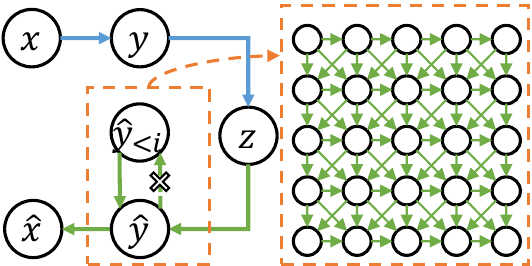}
             \label{fig:intro-ar}
         }
         \subfloat[Checkerboard Context\cite{He2021CheckerboardCM}]{
             \centering
             \includegraphics[width=.3\textwidth]{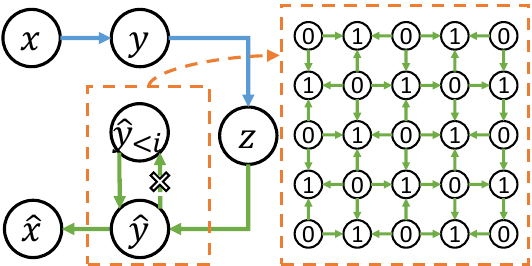}
             \label{fig:intro-arcb}
         }
    \end{minipage}
    
    \begin{minipage}[]{.3\textwidth}
         \subfloat[ELFIC (Data adaptive)\cite{Zhang2023ELFICAL}]{
             \centering
             \includegraphics[width=\textwidth]{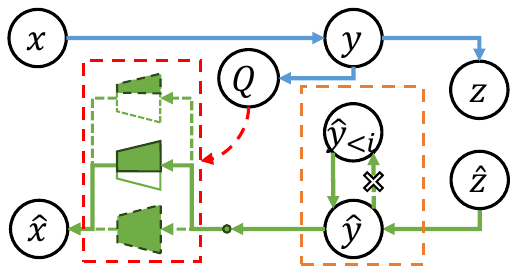}
             \label{fig:intro-slimda}
         }
         \vfill
         \subfloat[Cao2023 (Task adaptive)\cite{Cao2023SlimmableMI}]{
             \centering
             \includegraphics[width=\textwidth]{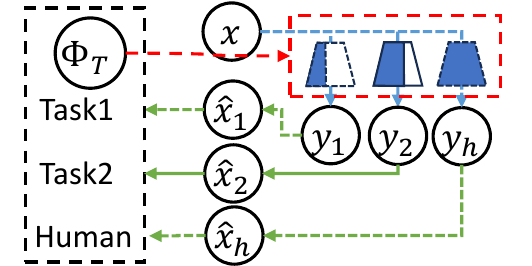}
             \label{fig:intro-slimta}
         }
    \end{minipage}
    \begin{minipage}[]{.6\textwidth}
         \subfloat[ABC (Ours)]{
             \centering
             \includegraphics[width=\textwidth]{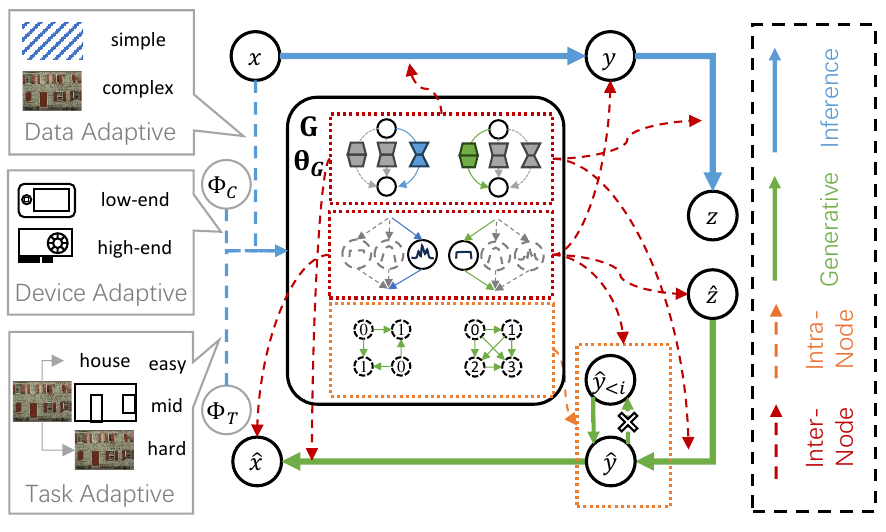}
             \label{fig:intro-ours}
         }
    \end{minipage}

    \caption{Bayesian network diagrams for various NIC frameworks. Our proposed framework ABC generalizes all concepts in (a-f).
    }
    \label{fig:intro}
\end{figure*}

\zyf{In order to explore adaptive computational scalability for NIC, we first re-examine existing NIC frameworks through the broader lens of the Bayesian network (BayesNet) paradigm.} \autoref{fig:intro} illustrates the BayesNet depiction of several examplary NIC architectures. We could see that all these methods are rooted in the hyperprior BayesNet framework (\autoref{fig:intro-hp}), which postulates that the generative model can be factorized as $p(\hat{\rvx}|\hat{\rvy})p(\hat{\rvy}|\hat{\rvz})p(\hat{\rvz})$\footnote{All symbols used in this paper are listed in Appendix \ref{sec:apd-notation}.}, where $\hat{\rvx}$ represents observed data, $\hat{\rvy}$ the prior, and $\hat{\rvz}$ the hyperprior. 

In this context, SlimCAE (\autoref{fig:intro-slim}) can be interpreted as allowing selective edge types between $\rvy$ and $\rvx$, and between $\hat{\rvx}$ and $\hat{\rvy}$, with each type manifesting different conditional dependencies, modeled via slimmable neural networks with varied channel widths. On the other hand, the joint autoregressive framework (\autoref{fig:intro-ar}) introduces autoregressive intra-node dependencies within $\hat{y}$ to enhance compression, at the expense of increased computational demands due to the requisite sequential processing\cite{He2021CheckerboardCM}. The checkerboard context framework (\autoref{fig:intro-arcb}) further improves such framework by designing parallelized intra-node dependencies for acceleration. 
Furthermore, existing NIC frameworks with adaptive computation control can also be represented through BayesNets with additional nodes: The ELFIC framework with data adaptive control (\autoref{fig:intro-slimda}) utilizes a quality list generated from $\hat{\rvy}$, which could be interpreted as a extra node $Q$ for selecting edge types between $\hat{\rvx}$ and $\hat{\rvy}$. The slimmable multi-task framework (\autoref{fig:intro-slimta}) basically selects a branch between $\rvy$ and $\rvx$ according to the target vision task, which could also be represented as an independent node $\Phi_{T}$.
These examples clearly demonstrate the intimate connection between the structure of the BayesNet and the complexity inherent in NIC frameworks.

Our proposed framework, as visualized in \autoref{fig:intro-ours}, introduces three key innovations in BayesNet structure. First, we model inter-node relationships using a heterogeneous bipartite BayesNet, designed to integrate seamlessly with dynamic neural networks like slimmable networks\cite{Yu2018SlimmableNN}. Second, we develop an intra-node BayesNet using multipartite or $k$-partite graphs, enabling parallel autoregressive processing whose speed can be controlled through the partition number $k$. Third, the framework achieves adaptive control by conditioning the BayesNet structure $\rmG$ on various factors. For example, we could apply $p(\rmG | \vx)$ for data adaptive control, $p(\rmG | \Phi_{C})$ for device adaptive control and $p(\rmG | \Phi_{T})$ for task adaptive control. Those conditional dependencies could be implemented together with an extra neural network module.

\subsection{Problem Definition for ABC}

Assume we have a set of observations $\rmX = \{ \rvx_{1}, \rvx_{2}, \ldots, \rvx_{N} \}$, the objective of BayesNet structure learning is to infer a distribution over the structure of the BayesNet $\rmG = (\rmV, \rmE, \vtheta)$ that best describes the observations. According to the Bayesian formalism, such optimization problem could be solved by minimizing expectation over $ p(\rmG | \rmX) $ of any function $f$ related to $\rmG$, under the constraint that $\rmG$ is acyclic:
\begin{equation}
\label{eq:bn-opt}
    \min_{\rmG} \E_{p(\rmG | \rmX)} [f(\rmG)], \quad  \mbox{s.t. } \rmG \mbox{ is acyclic}.
\end{equation}
For a computationally scalable compression framework, we tackle BayesNet structure learning, linking BayesNet configurations to computational complexity and shaping the function $f(\rmG)$ accordingly.

Apart from modeling of $\rmX$, the compression framework also involves latent variables $ \rmL = \{ \rvy_{1}, \rvy_{2}, \ldots, \rvy_{N}, \rvz_{1}, \rvz_{2}, \ldots, \rvz_{N}, \ldots \} $ to generate $\rmX$. 
Moreover, some independent nodes $\mPhi$ could be optionally deployed to control the generation of the BayesNet structure $\rmG$.
The resulting BayesNet should capture conditional dependencies within and between $\rmX $ and $ \rmL$. This generative relationship is described by:
\begin{equation}
\label{eq:bayes-gen-comp}
    p(\rmG, \rmL, \rmX, \mPhi) = p(\mPhi) p(\rmG | \mPhi) p( \rmL | \rmG ) p(\rmX | \rmG, \rmL).
\end{equation}
The expectation from \autoref{eq:bn-opt} is then: 
\begin{equation}
\label{eq:comp-exp}
    \E_{p(\rmG, \rmL | \rmX, \mPhi)} [f(\rmG, \rmL)] = \E_{p(\rmG | \rmX, \mPhi) p(\rmL | \rmX, \rmG)} [ f(\rmG, \rmL) ].
\end{equation}
Existing compression frameworks without $\rmG$ approximate $p(\rmL | \rmX)$ with an inference network $q(\rmL | \rmX)$, as in VAEs\cite{Kingma2013AutoEncodingVB}. Incorporating $\rmG$ into our design, we refine our inference network to $q(\rmG, \rmL | \rmX, \mPhi)$ for enhanced approximation: 
\begin{equation}
\label{eq:comp-exp-approx}
    \E_{p(\rmG, \rmL | \rmX, \mPhi)} [f(\rmG, \rmL)] \approx 
    \E_{q(\rmG | \rmX, \mPhi) q(\rmL | \rmX, \rmG)} [ f(\rmG, \rmL) ].
\end{equation}
This approach allows $\rmG$ to direct the computational complexity of both the inference and the generative processes. Furthermore, we adopt a differentiable approach to learning $\rmG$, in line with recent advances\cite{Yu2019DAGGNNDS, Lorch2021DiBSDB}, and incorporate this aspect into our model specification.

\subsection{Adaptive BayesNet Structure Generation}
The BayesNet structure $\rmG$ directly influences computational complexity, which can be adaptively controlled through the inference network $q(\rmG | \rmX, \mPhi)$. This network generates $\rmG$ based on both observed data $\rmX$ and controller nodes $\mPhi$. Here, we implement three types of adaptive control:

The \textbf{data adaptive controller} $q(\rmG | \rmX)$ could assign different complexity to the BayesNet according to the difficulty of compression and reconstruction of the observed data.
Naturally, this could be implemented with a shallow convolutional network.

The \textbf{device adaptive controller} $q(\rmG | \mPhi_{C})$ allows complexity adjustment by setting $\mPhi_{C}$ assigned to specific device computation budgets. In practice, this could be achieved by assuming $\mPhi_{C}$ as discrete distributions, which pre-defines a list of computation budgets of typical devices and optimizing the corresponding inference network $q(\rmG | \mPhi_{C})$ in batch:
\begin{equation}
    \label{eq:comp-opt-ada-dev}
    \begin{aligned}
    &\min_{\rmG} \E_{q(\rmG | \rmX, \mPhi) q(\rmL | \rmX, \rmG)} [ f(\rmG, \rmL) ] , \Phi_{C} \in \{1,2,3,\ldots\} \\
    &s.t. \forall \Phi_{C}, C(\hat{\rmG}^{\Phi_{C}}) \le C^{\Phi_{C}}
    \end{aligned}
\end{equation}
where $C(\hat{\rmG}^{\Phi_{C}})$ indicates the complexity of generated BayesNet sample from $q(\rmG | \mPhi_{C})$, and $C^{\Phi_{C}}$ indicates the computation budget assigned to the specific index of $\Phi_{C}$.

The \textbf{task adaptive controller} $q(\rmG | \mPhi_{T})$ enables automatic BayesNet control according to the required vision task for the reconstructed image. As discussed in \autoref{sec:rw-nic}, vision task adaption could be achieved by assigning specific modules to different tasks. This process could also be represented by altering BayesNet structures. Additionally, it is also possible to adjust computational complexity according to vision tasks.
Similar to $\mPhi_{C}$, $\mPhi_{T}$ could be also assigned to a discrete variable representing different vision tasks, which is usually optimized towards different distortion losses.

In some cases, it may be preferable to treat $\rmG$ as an independent generative node, by setting $q(\rmG | \rmX, \mPhi) = p(\rmG)$. This non-adaptive approach excludes $\rmG$ from the bitstream, requiring the same pre-defined $\rmG$ to be shared between compression and decompression processes.

\subsection{Optimizing Rate-Distortion-Complexity Trade-off}

The optimization objective of ABC is to obtain a codec with \textbf{lowest bit-rate $\Ls_{R}$}, \textbf{lowest reconstruction distortion $\Ls_{D}$} using \textbf{least computation resource $\Ls_{C}$}, under \textbf{certain adaptive control parameters $\hat{\mPhi}$}. Naturally, this can be formulated as:
\begin{equation}
    \label{eq:comp-opt}
    \begin{aligned}
    &\min_{\rmG} \Ls_{R}(\rmG, \rmL), \Ls_{D}(\rmG, \rmL), \Ls_{C} (\rmG, \rmL), \\
    &\mbox{s.t. } \rmG \mbox{ is acyclic}, \mPhi = \hat{\mPhi}.
    \end{aligned}
\end{equation}

Following standard lossy compression approaches\cite{Johnston2019ComputationallyEN, Yin2022ExploringSS}, we combine these objectives with weights $\lambda_{D}, \lambda_{C}$ controlled by $\mPhi$ to adjust their trade-off:
\begin{equation}
    \label{eq:comp-opt-combine}
    \begin{aligned}
    &\E_{q(\rmG | \rmX, \mPhi) q(\rmL | \rmX, \rmG)} [ f(\rmG, \rmL) ] = \\
    &\Ls_{R}(\rmG, \rmL) + \lambda_{D}(\mPhi) \Ls_{D}(\rmG, \rmL) + \lambda_{C}(\mPhi) \Ls_{C} (\rmG, \rmL).
    \end{aligned}
\end{equation}

Note that by setting $f(\rmG, \rmL)$ in line with the established rate-distortion (R-D) trade-off, \autoref{eq:comp-exp-approx} falls in step with loss functions prevalent in standard compression frameworks.

\noindent\textbf{Rate Loss $\Ls_{R}$: }
$\Ls_{R}$ could be computed by enacting the BayesNet's generative process with latent node samples $\hat{\rmL}$. However, in scenarios where complexity is tailored to input data or other controller nodes, the conditional inference model $q(\rmG | \rmX, \mPhi)$ should be employed. Note that in this case $\Ls_{R}$ should also include the bit-rate for samples $\hat{\rmG}$. This leads to:
\begin{equation}
    \label{eq:comp-opt-loss-rate}
    \begin{aligned}
    \Ls_{R}(\rmG, \rmL) =& \log_{2} p_{\rmG}(\hat{\rmG}) + \log_{2} p_{\rmL|\rmG}(\hat{\rmL} | \rmG), \\
    &\hat{\rmL} \sim q(\rmL | \rmX, \rmG), \hat{\rmG} \sim q(\rmG | \rmX, \mPhi).
    \end{aligned}
\end{equation}

\noindent\textbf{Distortion loss $\lambda_{D} \Ls_{D}$: }
$\Ls_{D}$ can be tailored to specific optimization objectives, which may encompass a range of complexity measures or their amalgamations, such as conventional metrics like Mean-Squared Error (MSE) or machine-vision oriented metrics like Frechet Inception Distance (FID)\cite{Heusel2017GANsTB}. 
As we may control the codec for a specific human or machine vision task, $\Ls_{D}$ could be defined as a weighted combination of multiple task-specific metrics, where the weights are affected by $\rmG$:
\begin{equation}
    \label{eq:comp-opt-loss-dist}
    \Ls_{D} (\rmG, \rmL) = \lambda_{T1} (\rmG) \Ls_{D1} (\rmG, \rmL) + \lambda_{T2} (\rmG) \Ls_{D2} (\rmG, \rmL) + \ldots
\end{equation}

$\lambda_{D}$ controls the trade-off between bit-rate and distortion. Usually codecs may allow users to set their required reconstruction quality, and in our framework this could be achieved through a distortion controller node $\Phi_{D}$. A straightforward implementation is to set $\Phi_{D}$ as a discrete distribution, whose samples selects the corresponding $\lambda_{D}$ and $\rmG$:
\begin{equation}
    \label{eq:comp-opt-loss-lambdad}
    \lambda_{D}(\mPhi) = \lambda^{\Phi_{D}}_{D}, \hat{\rmG} = \hat{\rmG}^{\Phi_{D}}, \Phi_{D} \in \{1,2,3,\ldots\}.
\end{equation}
This implementation is also aligned with non-variable-rate NIC frameworks which trains multiple models with different $\lambda_{D}$, such as \cite{Ball2018VariationalIC}, \cite{Cheng2020LearnedIC}, etc.

\noindent\textbf{Computation Loss $\lambda_{C} \Ls_{C}$: }
$\Ls_{C}$ can also include various complexity metrics or their combinations, such as the number of Floating-point Operations (FLOPs), network parameters, time, or energy consumption. 
However, complexity metrics has wide variable ranges, which requires $\lambda_{C}$ to be carefully tuned to adjust the trade-off between compression performance and complexity. In practice, a ratio definition of $\Ls_{C}$ could be used, such as calculating the linear interpolation ratio between the minimum complexity BayesNet $\rmG_{min}$ and maximum complexity BayesNet $\rmG_{max}$ of the current BayesNet:
\begin{equation}
    \label{eq:comp-opt-loss-complex-interp}
    \Ls_{C} (\rmG, \rmL) = \frac{C(\rmG) - C(\rmG_{min})}{C(\rmG_{max}) - C(\rmG_{min})}.
\end{equation}

In this way, $\lambda_{C}$ could be defined between $+\infty$ and 0, to control the optimal complexity between minimum complexity and maximum complexity. A possible definition is:
\begin{equation}
    \label{eq:comp-opt-loss-lambdac}
    \lambda_{C}(\mPhi) = - \log_{2} \mPhi_{C}, \Phi_{C} \sim \uniform(0,1).
\end{equation}

\subsection{Two-Stage Stable Optimization of $ f(\rmG, \rmL) $}
\label{sec:overview-opt-2stage}
While the solution presented in \autoref{eq:comp-opt-combine} serves as a comprehensive approach for optimizing multiple objectives with trade-offs, its convergence is critically challenged by the large structural search space introduced by the variable BayesNet $\rmG$.
Although enabling fine-grained control of computational complexity, the diversity of architectures within $\rmG$ contributes high-variance noise, which destabilizes the optimization of the rate-distortion-complexity trade-off.

To address this challenge, we constrain the search to a finite set of BayesNet architectures sampled from a discrete distribution.
The architectures within this distribution are pre-optimized under computational constraints, ensuring they lie near the Pareto front of the rate-distortion-complexity trade-off. 
Inspired by neural architecture search\cite{Liu2018DARTSDA}, our two-stage optimizer first explores the enlarged search space to identify these Pareto-optimal structures. Afterwards, it fine-tunes the model within this narrowed subspace to stabilize the rate-distortion optimization. This approach explicitly addresses the challenges posed by the structural diversity of $\rmG$, transforming an intractable search problem into a controllable optimization problem.


Specifically, the first stage aims to obtain an optimal rate-distortion trade-off under any possible BayesNet structures.
This optimization step could be represented as:
\begin{equation}
    \label{eq:comp-opt-combine-stable-s1}
    \forall \rmG, \min (\Ls_{R}(\rmG, \rmL) + \lambda_{D}(\mPhi) \Ls_{D}(\rmG, \rmL))
\end{equation}
The second stage focuses on selecting a finite set of BayesNet structures that satisfy specific complexity constraints.
Naturally, we could optimize the $\Ls_{C}$ component together with $\Ls_{R}, \Ls_{D}$. Similar to \autoref{eq:comp-opt-loss-complex-interp}, we could define the rate-distortion loss as a linear interpolation ratio. The corresponding optimization problem is defined as:
\begin{equation}
    \label{eq:comp-opt-combine-stable-s2}
    \min (\Ls_{RD} (\rmG, \rmL) + \lambda_{C}(\mPhi) \Ls_{C} (\rmG, \rmL)),
\end{equation}
where
\begin{equation}
 \Ls_{RD} (\rmG, \rmL) = \frac{\Ls_{RD}(\rmG) - \Ls_{RD}(\rmG_{min})}{\Ls_{RD}(\rmG_{max}) - \Ls_{RD}(\rmG_{min})}
\end{equation}
It is important to note that this approach is suitable only for differentiable complexity measures, such as FLOPs and number of network parameters. For non-differentiable metrics, such as execution time, a greedy search algorithm (e.g., as described in \cite{Yu2019AutoSlimTO}) is necessary to pinpoint the optimal configurations for different complexity levels. Alternatively, we may first determine the optimal configurations via greedy search, and then treat $\mPhi_{C}$ as a discrete distribution that samples from these optimal points to further refine the rate-distortion trade-off.

\section{Structure Learning for Inter and Intra-Node BayesNets}
\label{sec:method}


As previously noted in \autoref{eq:bn-opt}, $\rmG$ should reflect a distribution over DAG structures. For edges between different node groups in compression models, such as $p(\rvx | \rvy)$ or $p(\rvy | \rvz)$, ensuring a DAG structure is straightforward since bipartite graphs are basically acyclic as long as all edges comes from one partite to the other; we call this component the \textbf{Inter-node BayesNet} ($\rmG_{inter}$). For conditional dependencies within node groups, like $p(\ervy_{i} | \ervy_{i-1})$ which also require DAG adherence, we refer to it as the \textbf{Intra-node BayesNet} ($\rmG_{intra}$). The forthcoming sections will delve into the methodology for learning both $\rmG_{inter}$ and $\rmG_{intra}$ as DAGs.

\subsection{Heterogeneous Bipartite Graphs for Inter-node BayesNet}
\label{sec:method-inter}

In this section, we specify $\rmG_{inter}$. Note that $\rmG_{inter}$ is a bipartite graph, and as long as all edges comes from one partite to the other, it will inherently satisfy the acyclic constraint.


Normally, we could define bipartite graphs as adjacency matrices between nodes, which requires that both partite contains mutually independent splitted groups from the original node.
Nonetheless, this disables the integration of normalization layers such as GDN\cite{Ball2015DensityMO}, which is often integral to compression models.
To this end, we offer an innovative parameterization for the bipartite graph, tailored for heterogeneous DAG structure learning within image compression frameworks. Rather than dividing nodes into independent groups, we maintain individual nodes in adjacent layers and diversify only the connecting edges. Each node pair is linked through multiple neural network edges of varying sizes and complexities, from which only one is selected. In this context, $\rmG_{inter}$ is characterized by $N$ categorical distributions: 
\begin{equation}
\label{eq:bn-latent-hetero-prior}
    p(\rmG^{\hat{\rvy},\hat{\rvz}}_{inter}) = \categorical(\pi_{\hat{\rvy},\hat{\rvz}}), \quad \pi_{\hat{\rvy},\hat{\rvz}} \in \R^{N}.
\end{equation}
Thus, the connection $p(\hat{\rvy} | \hat{\rvz}, \rmG^{\hat{\rvy},\hat{\rvz}}_{inter})$ can be expressed as: 
\begin{equation}
\label{eq:bn-latent-hetero}
    p(\hat{\rvy} | \hat{\rvz}, \rmG^{\hat{\rvy},\hat{\rvz}}_{inter}) = \sum^{N}_{n} \hat{\mG}^{\hat{\rvy},\hat{\rvz}}_{inter}[n] p_{n}(\hat{\rvy} | \hat{\rvz}).
\end{equation}
This heterogeneous DAG learning approach does not merely determine \textbf{whether} nodes are interdependent; it specifies \textbf{how} they are interconnected. Given the inherent dependencies among latent nodes in compression frameworks, this nuanced parameterization is particularly apt for representing inter-node BayesNets.

~Once $\rmG_{inter}$ is established, we proceed to construct the heterogeneous edges, represented as $p_{n}(\hat{\rvy} | \hat{\rvz})$, utilizing neural networks. Following\cite{Yang2021SlimmableCA}, we employ slimmable networks and select the $n$-th channel width configuration for each $p_{n}(\hat{\rvy} | \hat{\rvz})$ to implement computation scalability. This approach is memory-efficient as it does not necessitate additional memory overhead compared to a singular network implementation for $p(\hat{\rvy} | \hat{\rvz})$. A comprehensive example detailing this implementation is provided in \autoref{sec:impl} for further reference.

Other than computational scalability, the heterogeneous graph can also be deployed for task adaptivity. Following \cite{Li2024ImageCF}, we adopt the Spatial-Frequency Adaption (SFA) modules to implement edges for different vision tasks. Moreover, SFA modules could also be implemented with slimmable networks, achieving computation scalability along with task adaptivity.

~$\rmG_{inter}$ could be optimized by leveraging samples from $p(\rmG_{inter})$ as weights to blend the various edges, enabling end-to-end optimization via the Gumbel-softmax reparameterization trick\cite{Jang2016CategoricalRW}, eliminating the need for an additional loss function. The computational complexity associated with this optimization can be quantified using the same mixing weights:
\begin{equation}
\label{eq:bn-latent-hetero-complexity}
    C(\rmG^{\hat{\rvy},\hat{\rvz}}_{inter}) = \sum^{N}_{n} \hat{\mG}^{\hat{\rvy},\hat{\rvz}}_{inter}[n] C(p_{n}(\hat{\rvy} | \hat{\rvz})),
\end{equation}
where $C(p_{n}(\hat{\rvy} | \hat{\rvz}))$ represents the complexity measure for a single edge such as FLOPs.


~Our approach to parameterizing $p(\rmG_{inter})$ bears resemblance to Differentiable Architecture Search\cite{Liu2018DARTSDA} in both concept and optimization techniques. Both methods utilize continuous relaxation to combine outputs from diverse neural network modules. However, the underlying objectives differ. NAS focuses on developing neural networks with enhanced performance by refining the computational graph. In contrast, our objective is to construct optimized probabilistic graphs tailored for compression, with neural networks serving as approximators of the conditional probabilities.




\subsection{Homogeneous Multipartite Graph for Intra-node BayesNet}
\label{sec:method-intra}

\begin{table*}[t]
    \centering
    \caption{Statistics of different intra-node BayesNet configurations for a $3 \times 32 \times 32$ tensor. Numbers inside the nodes represent the topological indices of the partites. For Maskconv each partite only include one node, so we omit the topological index.
    }
    \resizebox{\linewidth}{!}{
    \begin{tabular}{c|ccccc}
         & MaskConv\cite{Minnen2018JointAA} &  Zigzag\cite{Li2019EfficientAE} & Checkerboard\cite{He2021CheckerboardCM} &  Multistage\cite{Lin2023MultistageSC} & Channel\cite{Minnen2020ChannelWiseAE}\\
    & \includegraphics[width=.15\textwidth]{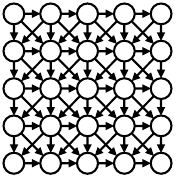}
    & \includegraphics[width=.15\textwidth]{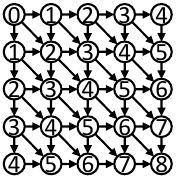}
    & \includegraphics[width=.15\textwidth]{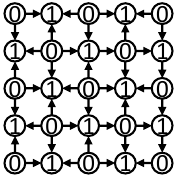}
    & \includegraphics[width=.15\textwidth]{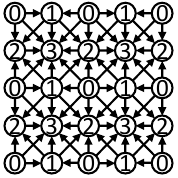}
    & \includegraphics[width=.15\textwidth]{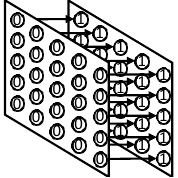} \\
    \hline
    Stages & 1024 & 63 & 2 & 4 & 3 \\
    FLOPs & 11718 & 8835 & 2976 & 11718 & 2048 \\
    Time (ms) & 11783.0  & 727.3  & 46.5  & 82.6  &  71.2 
    \end{tabular}
    }
    \label{tab:ar-stat}
\end{table*}

As noted in \autoref{sec:rw-bnsl}, traditional BayesNet structure learning has been constrained by the need to maintain a DAG, which, when applied to image data with its vast number of pixels, can greatly reduce efficiency. \autoref{tab:ar-stat} illuminates an alternative approach that can satisfy both acyclic and parallel constraints effectively. It reveals that the number of parallel stages corresponds to the multipartite quantity within each BayesNet. Since edges can only direct from topologically earlier nodes to later ones, nodes within the same partite can be processed simultaneously. Bayesian inference can thus be executed by sequentially processing each partite from the first to the last. This process not only maintains acyclicity but also confines the number of parallel stages to the number of partites. Hence, we propose to parameterize $p(\rmG_{intra})$ with categorical distributions at each node, each category corresponding to the node's topological index within the partites. Consequently, $p(\rmG_{intra})$ takes the following form:
\begin{equation}
\label{eq:ga-dist}
    p(\rmG_{intra}) = \prod^{C,H,W}_{c,h,w} p(\ermT_{c,h,w}), \quad \ermT_{c,h,w} = \categorical(\pi_{c,h,w}),
\end{equation}
where $\ermT_{c,h,w}$ denotes the topological indices of partites per categorical distribution, with $C,H,W$ being the node count along the channel, height, and width dimensions, respectively. Adjusting the number of partites (or parallel stages) $\dim \pi_{c,h,w}=S_{intra}$ can modulate the computation time.

Having constrained the number of partites to manage computational time, we now focus on learning the intra-node BayesNet structure $\rmG_{intra}$ to minimize bit-rate. The optimization challenge is formulated as:
\begin{equation}
\label{eq:ga-opt}
    \min_{p(\rmG_{intra})} \E_{p(\rmG_{intra})} p(\rmX, \rmL | \rmG_{intra}), \quad \mbox{s.t. } \hat{\rmG}_{intra} \mbox{ is acyclic},
\end{equation}
where $\hat{\rmG}_{intra}$ is sampled from $p(\rmG_{intra})$. However, unlike the optimization for $\rmG_{inter}$, partite indices are ordinal and don't lend themselves to being used as mixing weights, making the Gumbel-softmax reparameterization trick inapplicable. We therefore approach $\rmG_{intra}$ as discrete variables and employ the Monte-Carlo method with the VIMCO objective\cite{Mnih2016VariationalIF} for optimization:
\begin{equation}
\label{eq:ga-opt-loss}
    \Ls_{\rmG_{intra}} = \E_{q(\rmG_{intra}^{1:M} | \rmX)} \log \frac{1}{M} \sum^{M}_{i=1} \log p(\rmX, \rmL | \rmG_{intra}^{i}),
\end{equation}
where $\rmG_{intra}^{i}$ is the i-th Monte-Carlo sample and $M$ denotes number of samples.

Moreover, note that \autoref{eq:ga-dist} presumes all partites are independent—an unlikely scenario as proximal nodes often exhibit stronger correlations. In practice, we can approximate $p(\rmG_{intra})$ with an unconditional generative model to consider such correlations and optimize this proxy instead, described as:
\begin{equation}
\label{eq:ga-dist-approx}
    p(\rmG_{intra}) \propto g(\rmN), \quad \rmN \sim \mathcal{N}(0,1),
\end{equation}
with $ g(\rmN) $ symbolizing a generative network derived from standard normal noise.

~With the model $p(\rmG_{intra})$ established, we can draw samples of topological indices for partites, denoted $\hat{\emT}_{c,h,w}$, from each respective $\ermT_{c,h,w}$. The subsequent step involves establishing conditional dependencies among these groups. Since dependencies are restricted to nodes from different partites, a dynamic masked convolution operation is utilized. This operation, which could be called multipartite-based dynamic masked convolution, ensures each node's dependencies are limited to preceding nodes within the convolutional kernel's range. For instance, the intra-node BayesNet for $\hat{\rvy}$ can be characterized by:
\begin{equation}
\label{eq:impl-ar-dynconv}
    \prod^{C,H,W}_{c,h,w} p(\ervy_{c,h,w} | \rmG_{intra}) \propto \dynconv(\hat{\ervy}_{c,h,w}, \etW_{c,h,w}(\hat{\emT}_{c,h,w})), 
\end{equation}
where $\hat{\ervy}_{c,h,w}$ are samples from $\ervy_{c,h,w}$, and $\etW_{c,h,w}$ represents a dynamically adjusted convolutional kernel influenced by adjacent nodes' topological indices. As an illustration, consider a 2D $5 \times 5$ convolution: the dynamic kernel $\etW_{c,h,w}$ used for node $\ervy_{c,h,w}$ is defined by:
\begin{equation}
\label{eq:impl-ar-maskconv}
    \begin{aligned}
    &\etW_{c,h,w}[i,j+2,k+2] = \Bigg\{
    \begin{aligned}
        &\etW_{i,j+2,k+2}&, \hat{\emT}_{c,h,w} < \hat{\emT}_{i,h+j,w+k} \\
        &0&, \hat{\emT}_{c,h,w} \ge \hat{\emT}_{i,h+j,w+k}
    \end{aligned}
    , \\
    &i \in \{0,1,\ldots,C-1\}, j,k \in \{-2,-1,0,1,2\}.
    \end{aligned}
\end{equation}
For in-depth guidance of its implementation, please refer to Appendix \ref{sec:apd-impl}.

\section{From BayesNet to ABC}
\label{sec:impl}

\begin{figure*}[t]
    \centering
    \includegraphics[width=.9\textwidth]{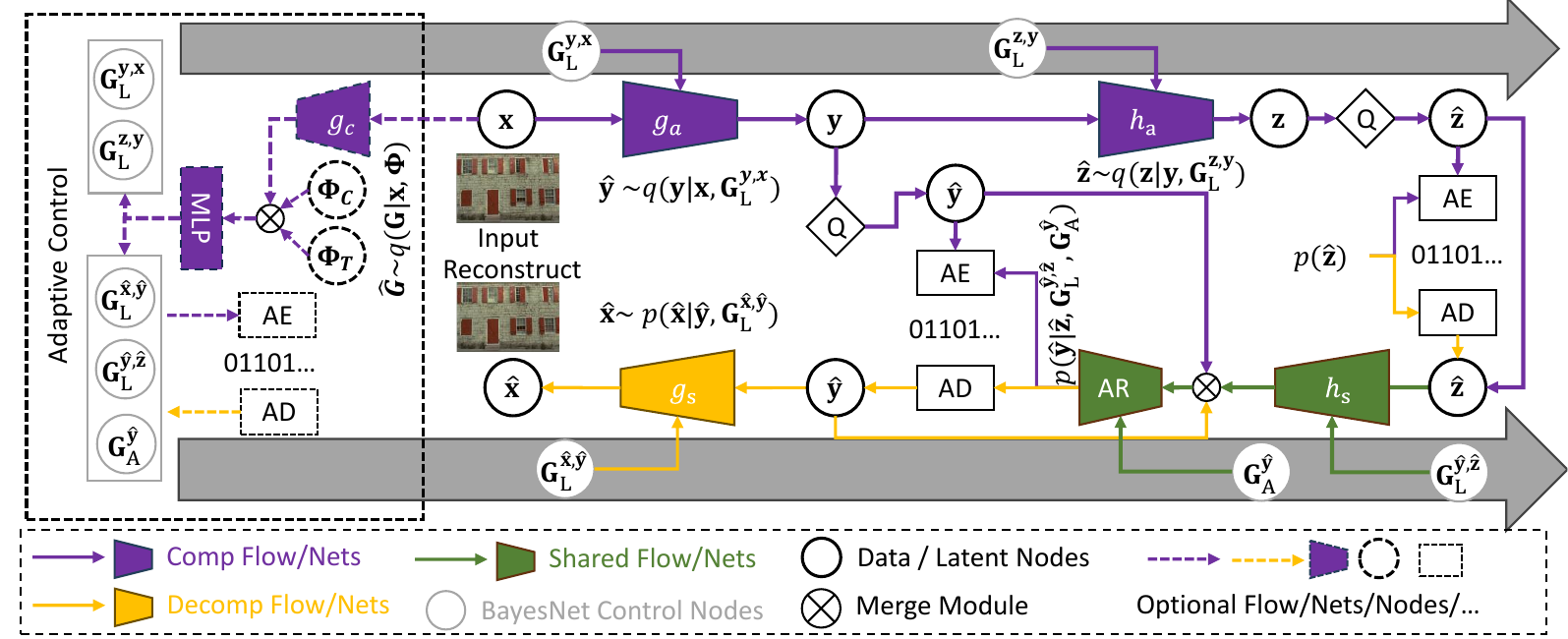}
    \caption{The architecture of the implemented example based on the joint autoregressive framework\cite{Minnen2018JointAA}. Comp/Decomp means Compression / Decompression. AE/AD symbolize the arithmetic encoder and decoder, respectively. AR denotes the autoregressive model that executes the learned intra-node BayesNet. Q stands for quantization. $\vx$ correspond to input image and $\hat{\vx}$ the reconstructed image. 
    }
    \label{fig:impl-example}
\end{figure*}

Upon determining the BayesNet structure, we proceed to specify the nodes and edges in this BayesNet, crafting a novel NIC framework, ABC. This section details an implementation instance of the envisioned computationally scalable NIC, leveraging the established joint autoregressive and hyperprior compression model\cite{Minnen2018JointAA}. It should be noted that this represents just one approach to implementing ABC, and alternative strategies may yield even better results. The architecture of our example is depicted in \autoref{fig:impl-example}, and we will explore how it seamlessly integrates with the aforementioned BayesNets.\footnote{Further details are provided in Appendix \ref{sec:apd-impl}.}


\subsection{Inter-node BayesNet to Backbone networks}
~The inter-node BayesNet is constructed over each edge between nodes within the original BayesNet, specifically:
\begin{equation}
    \begin{aligned}
    &q(\rvy | \rvx, \rmG^{\rvy,\rvx}_{inter}) \propto g_{a}(\rvx, \rmG^{\rvy,\rvx}_{inter}), \\
    &q(\rvz | \rvy, \rmG^{\rvz,\rvy}_{inter}) \propto h_{a}(\rvy, \rmG^{\rvz,\rvy}_{inter}), \\
    &p(\hat{\rvy} | \hat{\rvz}, \rmG^{\hat{\rvy},\hat{\rvz}}_{inter}) \propto h_{s}(\hat{\rvz}, \rmG^{\hat{\rvy},\hat{\rvz}}_{inter}), \\
    &p(\hat{\rvx} | \hat{\rvy}, \rmG^{\hat{\rvx},\hat{\rvy}}_{inter}) \propto g_{s}(\hat{\rvy}, \rmG^{\hat{\rvx},\hat{\rvy}}_{inter})
    \end{aligned}
\end{equation}
To achieve computational scalability, heterogeneous bipartite graphs are employed as outlined in \autoref{eq:bn-latent-hetero} for $\rmG^{\rvy,\rvx}_{inter}, \rmG^{\rvz,\rvy}_{inter}, \rmG^{\hat{\rvy},\hat{\rvz}}_{inter}, $ and $\rmG^{\hat{\rvx},\hat{\rvy}}_{inter}$. The implementation of $g_{a}, g_{s}, h_{a}$, and $ h_{s}$ follow the slimmable network in~\cite{Yang2021SlimmableCA}, with dynamic channel widths including 48, 72, 96, 144, and 192 for computational scalability. Note that we use a fixed 192 channels for all latent variables $\rvy, \rvz$, and only channels in the middle layers of each slimmable network are variable. This modification decouples bitrate scalability from computational scalability, enhancing the convenience in real-world applications.
Optionally, to achieve task adaptability combined with computational scalability, $\rmG^{\rvy,\rvx}_{inter} $ and $\rmG^{\hat{\rvx},\hat{\rvy}}_{inter}$ could be further extended to support multiple branches of SFA modules. Following \cite{Li2024ImageCF}, we adopt half of the corresponding dynamic channel widths as the number of hidden channels of SFA, which is 24, 36, 48, 72, and 96 for slimmable convolutional layers.

In practice, by modulating all $\rmG_{inter}$, which govern the channel width in slimmable networks, we can fine-tune the computational load of the inter-node neural networks.


\subsection{Intra-node BayesNet to Autoregressive models}
In the established framework, the intra-node BayesNet employs masked convolution applied to $\hat{\rvy}$, which forms the autoregressive model in NIC. We substitute this with our suggested homogeneous multipartite graph-based intra-node BayesNet $p(\hat{\rvy} | \rmG^{\hat{\rvy}}_{intra})$, which enables parallel computing on GPUs. The intra-node BayesNet uses a $5 \times 5$ dynamic masked convolution as in \autoref{eq:impl-ar-maskconv}, where $C$ equals number of partites, and $H$ and $W$ equals image height and width, respectively. 
To facilitate the optimization of the intra-node BayesNet, it is essential to include an additional VIMCO loss, as expressed in \autoref{eq:ga-opt-loss}, in the overall loss function.

In practice, by varying the number of partites in $\rmG^{\hat{\rvy}}_{intra}$, we can adjust the stages of parallel computations in autoregressive models, thereby influencing the decompression time.

\subsection{Adaptive Control Module}
To establish adaptive control, we implement $q(\rmG | \rmX, \mPhi)$, which employs a specialized branch module to forecast an appropriate Bayesian network structure tailored to the computational budget, image content, or follow-up vision tasks. This branch module comprises a set of conditionally selected sub-networks, a merge module for integrating diverse control inputs, and a multilayer perceptron (MLP) tasked with generating $\rmG$. For data adaptive control, 
we utilize a VGG-based network $g_{c}$ to predict from input data $\vx$. For device and task adaptive control, samples from $\Phi_{C}$ and $\Phi_{T}$ are concatenated directly to inform the control mechanism.

Note that the generated samples $\hat{\rmG}$ are processed in two different ways. 
During the compression phase, the structures of the inference Bayesian network, namely $\rmG^{\rvy,\rvx}_{inter}$ and $\rmG^{\rvz,\rvy}_{inter}$, can be omitted from the bitstream since they are only needed for the inference process. Conversely, the structures required for the generative process must be included in the compressed data for use during decompression. These generative structures consist of simple discrete variables that require minimal storage; thus, their contribution to the overall bit-rate is negligible compared to other components (from $\hat{y}$ and $\hat{z}$) that dominate the bitstream composition. As a result, these variables do not factor into the bit-rate optimization process.

\subsection{Optimization}
As described in \autoref{sec:overview-opt-2stage}, the optimization process involves two stages. The first stage only optimizes the rate-distortion loss given a randomly sampled BayesNet structure, as in \autoref{eq:comp-opt-combine-stable-s1}. The second stage optimize a loss function that simultaneously accounts for both the BayesNet structure (\autoref{eq:ga-opt-loss}) and the neural network parameters (\autoref{eq:comp-opt-combine-stable-s2}):
\begin{equation}
    \label{eq:impl-loss}
        \Ls = \Ls_{RD} + \lambda_{C}(\mPhi) \Ls_{C} + \Ls_{\rmG_{intra}}
\end{equation}

For the distortion loss during training, we employ mean squared error (MSE) loss in the first stage of optimization. In the second stage, we use MSE, MS-SSIM, or vision-task distortion loss to develop models for PSNR, MS-SSIM and machine vision tasks, respectively.
For the distortion loss associated with machine vision tasks, we follow \cite{Chen2023TransTICTT}, utilizing perceptual loss based on a pretrained ResNet50 from \textit{torchvision} (for classification) and Mask-RCNN from \textit{detectron2} (for object detection and instance segmentation).

\section{Experiments}
\label{sec:exp}

\subsection{Experiment Setup}

\textbf{Experiment Tasks}
~Following prior research\cite{Yang2021SlimmableCA}, our experiments focus on the task of lossy image compression with an emphasis on computational scalability. 
In our experiments, we evaluate the proposed ABC framework against a range of codecs that offer computational scalability, analyzing both their compression efficiency and the associated computational demands.
We extend our evaluation to include the performance of these scalable codecs in both human and machine vision tasks.
Furthermore, we test our codec's performance in lossy image compression against both computational scalable and non-scalable neural image compression (NIC) codecs under fixed computational limits, which is a common requirement for practical applications.
Finally, we validate the efficacy of the proposed intra-node BayesNet learning method, as well as conducting ablation studies to compare different implementations of the proposed ABC framework.
Further experimental details and results can be found in Appendix \ref{sec:apd-impl-trainval} and \ref{sec:apd-exp}, respectively.

\textbf{Metrics}
To quantify compression performance, we compute Bits Per Pixel (BPP) from the compressed bitstream. For human perceptual quality assessment, we employ established metrics: Peak Signal-to-Noise Ratio (PSNR) and Multi-Scale Structural Similarity Index (MS-SSIM). For machine vision tasks, we measure task-specific performance using top-1 accuracy (Acc@1) for image classification and mean Average Precision (mAP) for instance segmentation.
To holistically evaluate performance across bit-rate levels, we adopt the Bjontegaard-Delta Rate (-BD-Rate) metric. This calculates the average bitrate reduction percentage relative to a baseline method—the static-computation Hyperprior codec\cite{Ball2018VariationalIC} in our case—while preserving equivalent reconstruction quality. When applied to machine vision tasks, -BD-Rate measures bitrate savings under identical task accuracy levels.

\textbf{Datasets} 
~We use a merged image dataset from ImageNet\cite{Deng2009ImageNetAL} and CLIC\cite{CLIC} for training the whole framework, preprocessed following the method in\cite{He2022ELICEL}. The Kodak\cite{Kodak} and CLIC 2020 validation\cite{CLIC} dataset serves for both validation and testing.
Moreover, ImageNet\cite{Deng2009ImageNetAL} and MS-COCO\cite{Lin2014MicrosoftCC} are used in testing experiments for machine vision tasks.

\textbf{Implementation Details} 
~
Our framework builds upon the joint autoregressive architecture\cite{Minnen2018JointAA} outlined in \autoref{sec:impl}. 
Unless otherwise specified, we follow CompressAI\cite{begaint2020compressai} to set the hyperparameter $\lambda_{D}$ to optimize PSNR and MS-SSIM models under 4 different bit-rate levels, and follow $\lambda_{D}$ in \cite{Chen2023TransTICTT} for vision-oriented models.
The computational complexity parameter $\lambda_{C}$ is sampled from the discrete set $[0.0, 0.25, 0.5, 1.0, 2.0, 4.0, 8.0, 16.0] $, during second-stage optimization, yielding eight distinct complexity levels. These are designated as "ABC-L0" (maximum complexity) through "ABC-L7" (minimum complexity), hereafter referred to as "ABC-max" and "ABC-min" respectively in comparative analyses.
Note that to ensure equitable comparison of computational time across experiments, all tests are conducted on the identical hardware with an Intel i7-6800K CPU, NVIDIA RTX 2080Ti GPU, and 32GB of RAM.
\footnote{Further implementation details are included in \autoref{sec:apd-impl} in the Appendix.}

\subsection{Comparison to Other Computation Scalable Frameworks}
\label{sec:exp-sc}

This section assesses our computational scalable ABC framework against other computational scalable frameworks. In our experimental evaluation, we concentrate on two principal metrics to assess computational complexity: (1) the total Multiply-Add Computation (MAC), a commonly accepted indicator of neural network computational load, and (2) the total processing speed (including compression and decompression), which is crucial in real-world applications.

\subsubsection{Comparison under MAC computation metric}
\label{sec:exp-sc-mac}
\begin{figure*}[t]
\centering

    \subfloat[PSNR]{
        \centering
        \includegraphics[width=.4\linewidth]{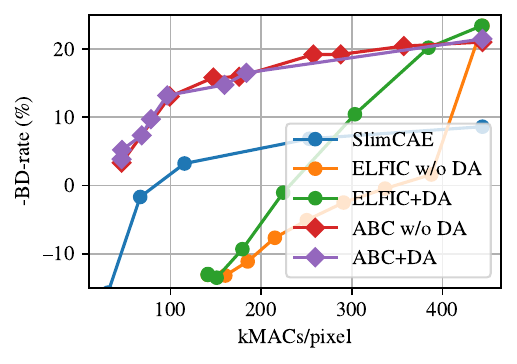}
    }    
    \subfloat[MS-SSIM]{
        \centering
        \includegraphics[width=.4\linewidth]{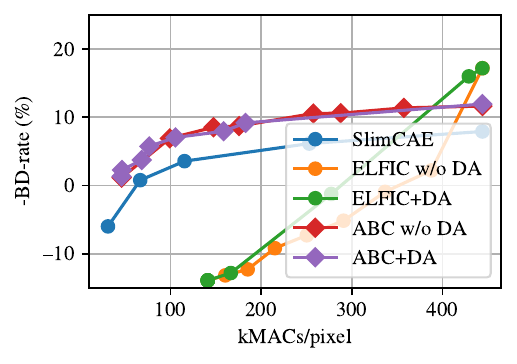}
    }   
    \caption{Comparative results for different computational scalable frameworks under MAC computation metric on Kodak dataset.}
    \label{fig:exp-abl-latent-kodak}

\end{figure*}

\begin{figure*}[t]
\centering
    \subfloat[PSNR]{
        \centering
        \includegraphics[width=.4\linewidth]{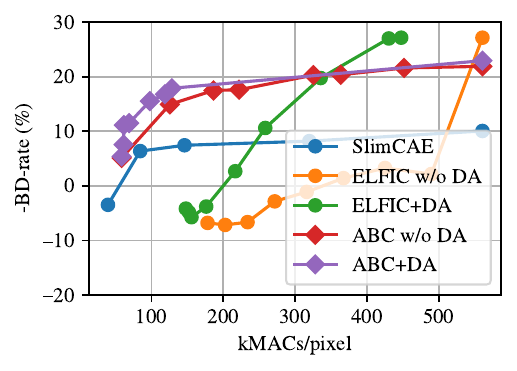}
    }    
    \subfloat[MS-SSIM]{
        \centering
        \includegraphics[width=.4\linewidth]{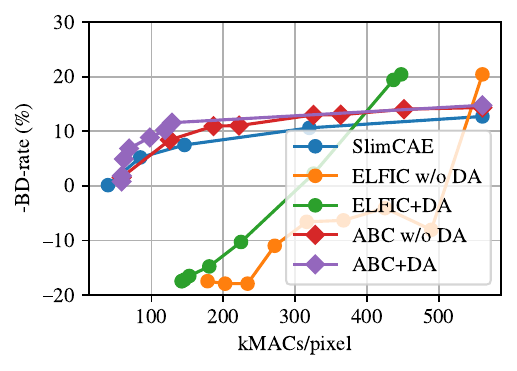}
    }     
    \caption{Comparative results for different computational scalable frameworks under MAC computation metric on CLIC dataset.}
    \label{fig:exp-abl-latent-clic}

\end{figure*}

As MAC is basically a neural network computation metric, we only compare with other computational scalable NIC codecs including SlimCAE\cite{Yang2021SlimmableCA} and ELFIC\cite{Zhang2023ELFICAL}, and exclude traditional codecs in this section. 
We report BD-Rate savings under diverse MACs for all methods.
For fair comparison, we adopt the same backbone network as ours, which is based on the Joint Autoregressive\cite{Minnen2018JointAA} framework) for all frameworks. 
Moreover, the autoregressive modules are also same as \cite{Minnen2018JointAA} across all frameworks as they have fixed MAC.
Note that $\lambda$-scheduling is deactivated in SlimCAE to facilitate uniform BD-rate analysis across levels, and slimmable context model is used as the autoregressive model according to the original paper. 
For ELFIC we also apply the same backbone networks and autoregressive models for fair comparison, and their major difference to ours is the universal slimmable network deployed for the synthesis transform network $g_{s}$.
Additionally, as ELFIC and our codec both support data-adaptive computational scalability, we include both results of the variant with and without data-adaptivity of both codecs (suffixed with "+DA" and "w/o DA" respectively).
The results are depicted in \autoref{fig:exp-abl-latent-kodak} for Kodak dataset and \autoref{fig:exp-abl-latent-clic} for CLIC dataset.

\begin{figure*}[t]
\centering
    \subfloat[Acc@1 (Classification)]{
        \centering
        \includegraphics[width=.4\linewidth]{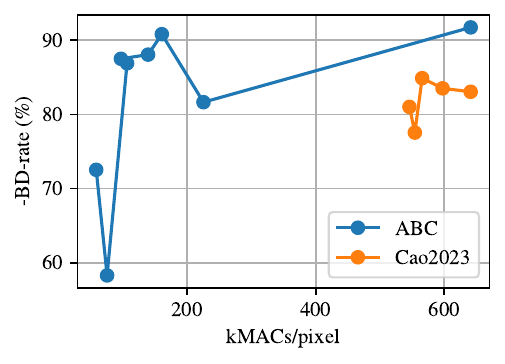}
    }
    \subfloat[mAP (Segmentation)]{
        \centering
        \includegraphics[width=.4\linewidth]{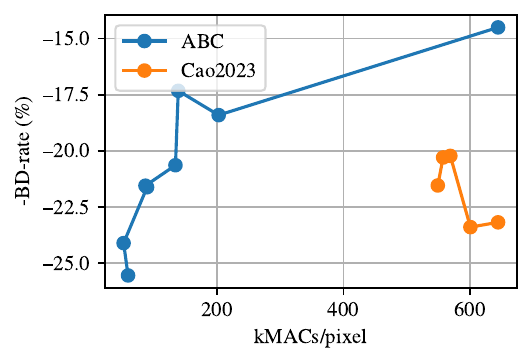}
    } 
    \caption{Comparative results for different computational scalable frameworks under MAC computation metric for image classification on ImageNet dataset and instance segmentation on MS-COCO dataset.}
    \label{fig:exp-abl-latent-mt}

\end{figure*}

Furthermore, we apply our proposed codec to various machine vision tasks and compare its performance with the recently introduced Cao2023 codec\cite{Cao2023SlimmableMI}, which is computationally scalable and designed for multi-task neural image compression (NIC). Specifically, we focus on the tasks of image classification and object instance segmentation.
To ensure a fair comparison, we implement backbone networks using SFA modules for both our codec and Cao2023. The results are shown in \autoref{fig:exp-abl-latent-mt}.

\begin{figure}[t]
\centering

    \includegraphics[width=.8\linewidth]{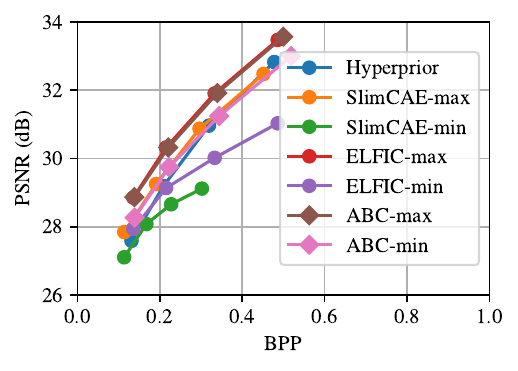}
    \caption{Rate/Distortion plots for different computational scalable frameworks under various computation levels on Kodak dataset.}
    \label{fig:exp-abl-latent-kodak-rd}

\end{figure}

For non-data-adaptive frameworks, first, our method demonstrates performance improvement over SlimCAE at all complexity levels, likely attributable to the fixed channels of latent variables that enhance performance under higher bit-rate. 
While SlimCAE\cite{Yang2021SlimmableCA} employs $\lambda$-scheduling to tackle this, it inadvertently ties bit-rate and distortion to computational complexity, which is less desirable for cases like low-powered devices needing high-quality images. In contrast, our inter-node BayesNet independently modulates the computational complexity of individual neural network modules, as evidenced by the rate-distortion plots in \autoref{fig:exp-abl-latent-kodak-rd} that show minimal impact on performance with reduced computation.
Furthermore, our method enables finer-grained complexity control by assigning layer-specific channel widths via distinct BayesNets, whereas SlimCAE enforces uniform channel widths across all layers, limiting adaptability.

When compared to non-data-adaptive ELFIC, our framework achieves marginally better performance at lower complexity levels, though slightly underperforming at maximum complexity. This discrepancy arises from ELFIC’s "sandwich rule," which prioritizes optimization for the largest model. Additionally, ELFIC’s computational scalability is restricted to the synthesis network, resulting in a narrower complexity range. Similar trends emerge in comparisons with data-adaptive ELFIC variants, where our method maintains superior overall performance with broader scalability.

Evaluating our data-adaptive codec against its non-adaptive counterpart reveals comparable compression performance across most complexity levels. Notably, on the CLIC dataset under low MAC budgets, data adaptivity improves results by tailoring different computational resources to individual images. However, the added computational overhead from the adaptive control module leads to marginally reduced performance at specific complexity thresholds.
These observations highlight that while data adaptivity offers conditional advantages, further refinement of adaptive control mechanisms is preferred to optimize the balance between performance and computational efficiency.



The multi-task compression results, as demonstrated in \autoref{fig:exp-abl-latent-mt}, highlight that our framework achieves marginally superior compression performance while supporting a substantially wider spectrum of computational complexities compared to existing methods. This capability originates from our full-module computation control mechanism, which enables task-specific adaptation across the entire architecture. In contrast, Cao2023 restrict slimmable network designs to the analysis network, limiting flexibility.
Notably, unlike human-centric quality metrics, vision task performance does not exhibit a strict positive correlation with MACs. This observation suggests that current multi-task compression implementations incur inherent architectural overhead, which is a key target for optimization in future studies.

Collectively, our framework demonstrates enhanced adaptability and efficiency in balancing computational demands across heterogeneous vision tasks, while establishing a foundation for advancing computational scalable multi-task NIC architectures.


\subsubsection{Comparison under total processing speed}
\label{sec:exp-sc-time}
This section evaluates our codec against traditional scalable codecs WebP\cite{libwebp} and BPG(444)\cite{libbpg} as well as formerly compared NIC codecs by evaluating total processing speed on Kodak images.
Data-adaptive codecs are excluded from this analysis, as time consumptions are non-differentiable and thus incompatible with our optimization pipeline, as detailed in \autoref{sec:overview-opt-2stage}.
Note that as autoregressive models dominate computational overhead in practical image coding, and former NIC codecs have no scalable design for autoregressive models, we implement a Hyperprior-based (non-autoregressive) framework for SlimCAE and ELFIC to compare at similar level of speed.



\begin{figure}[t]
\centering
    \includegraphics[width=.8\linewidth]{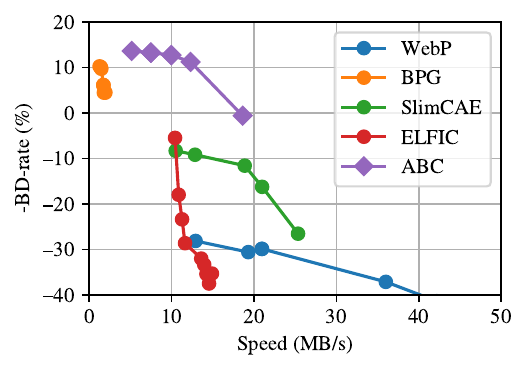}
    \caption{Comparison for computational scalable frameworks subject to total processing speed.}
    \label{fig:exp-sc-speed}
    \vspace{-6pt}
\end{figure}

The results illustrated in Figure \autoref{fig:exp-sc-speed}, reveal three key findings. First, our codec achieves rate-distortion parity with BPG and more than 30\% higher compression performance than BPG, by leveraging GPU parallelism to improve encoding latency.
Second, by integrating a scalable intra-node BayesNet as its autoregressive core, our framework surpasses existing scalable NIC baselines in both adaptability and performance.
Finally, our codec supports a much wider computational adjustment range than ELFIC. This stems from ELFIC’s GPU-accelerated synthesis network contributing minimally (<30\%) to total latency, whereas our scalable autoregressive module governs more than 70\% of runtime—enabling finer-grained complexity control.

\subsubsection{Qualitative Results}
To empirically validate computational scalability, we present a systematic visual comparison of reconstructed images across varying complexity levels in \autoref{fig:exp-sc-vis}. The evaluation includes methods benchmarked in earlier experiments (ABC-CLS denotes our proposed classification-optimized framework), with each method visualized at three operational modes—minimum, intermediate, and maximum complexity — maintaining a target BPP of approximately 0.3.

Qualitative inspection reveals negligible perceptual degradation across complexity levels for most methods, even when comparing the minimum and maximum modes. However, detailed scrutiny of high-frequency regions exposes discernible artifacts (e.g., texture blurring and quantization noise) in low-complexity configurations. This observation aligns with theoretical expectations, as reduced network capacity inherently limits representational capability, particularly affecting fine-grained feature preservation.
These visual findings complement our quantitative metrics, demonstrating that while our framework maintains perceptual consistency across computational budgets, architectural constraints in lightweight implementations inevitably introduce fidelity compromises — a critical consideration for edge deployment scenarios.

\begin{table*}
    \centering
    \caption{Visualization of Reconstructed Images using computational scalable codecs under different computation levels.}
    \newcommand{\ImageWidthVariable}{5cm}
    \resizebox{\linewidth}{!}{
    \begin{tabular}{c|ccc}
    \toprule
    Methods & Min & $\longrightarrow$ & Max \\
    \midrule
    & \includegraphics[width=\ImageWidthVariable]{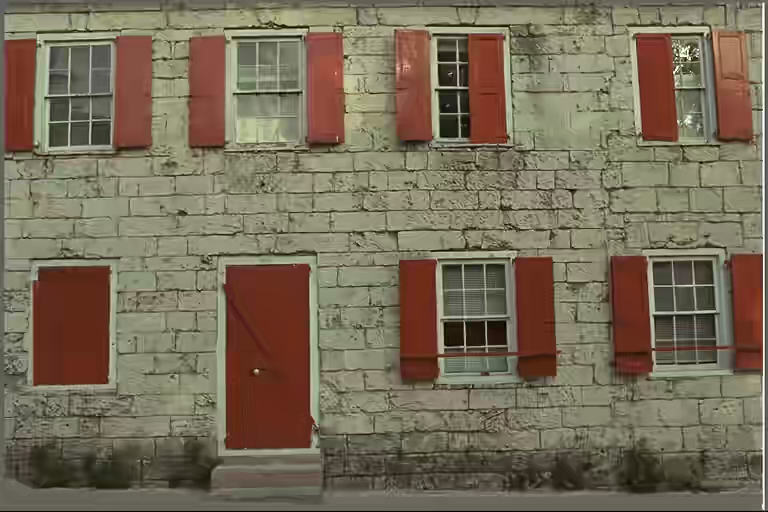} & \includegraphics[width=\ImageWidthVariable]{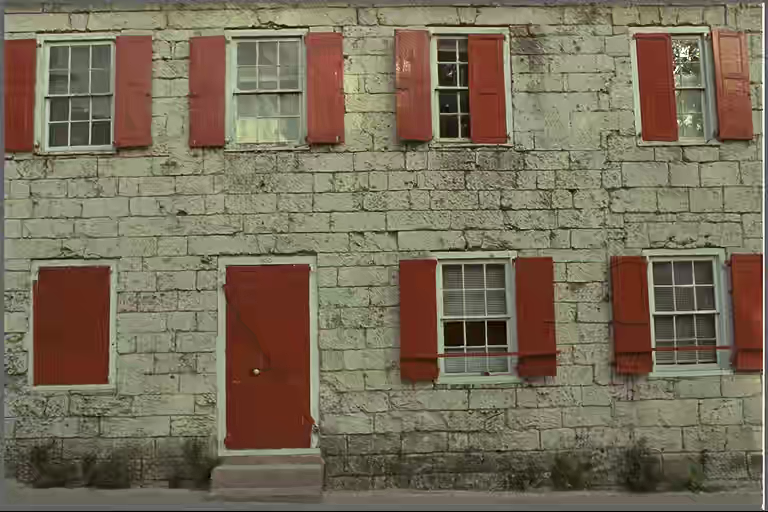} & \includegraphics[width=\ImageWidthVariable]{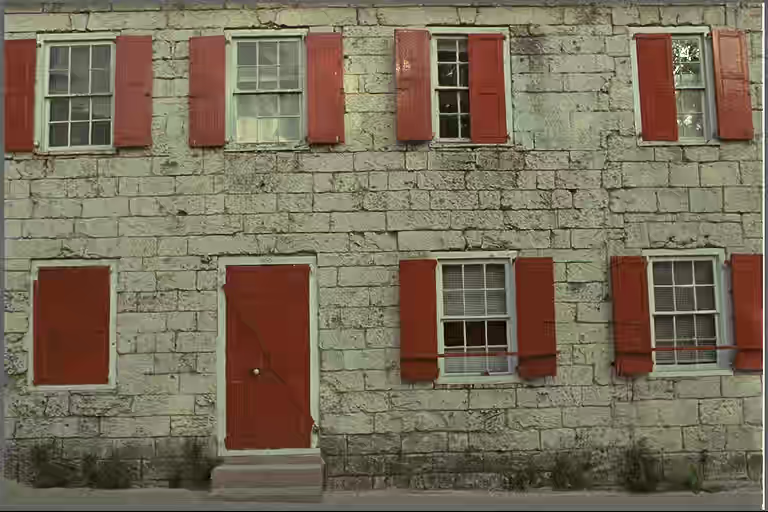} \\
    BPG & BPP=0.28,PSNR=26.65 & BPP=0.38,PSNR=27.47 & BPP=0.38,PSNR=27.39 \\
    & \includegraphics[width=\ImageWidthVariable]{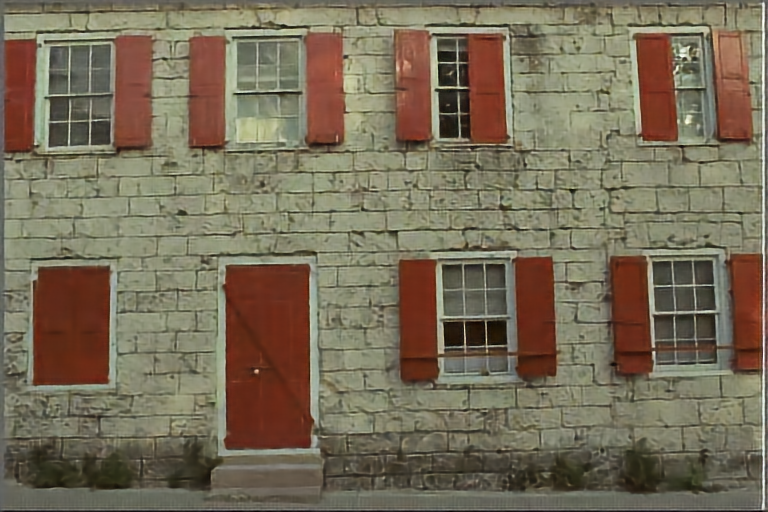} & \includegraphics[width=\ImageWidthVariable]{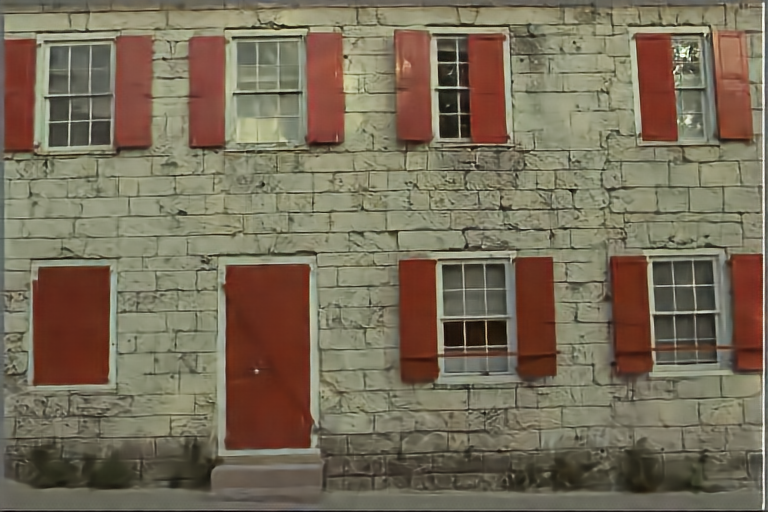} & \includegraphics[width=\ImageWidthVariable]{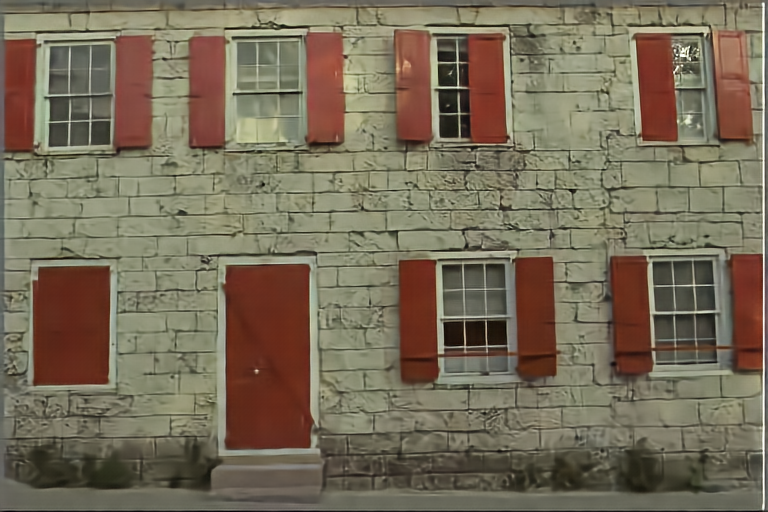} \\
    SlimCAE & BPP=0.34,PSNR=26.29 & BPP=0.31,PSNR=26.84 & BPP=0.30,PSNR=26.85 \\
    & \includegraphics[width=\ImageWidthVariable]{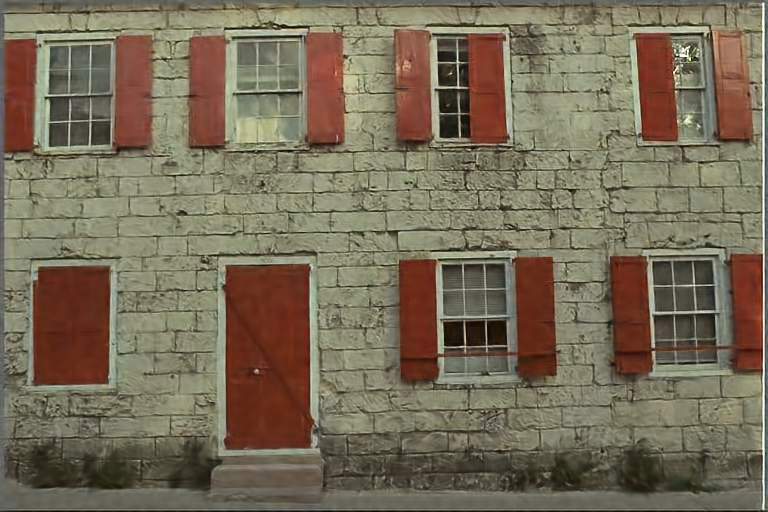} & \includegraphics[width=\ImageWidthVariable]{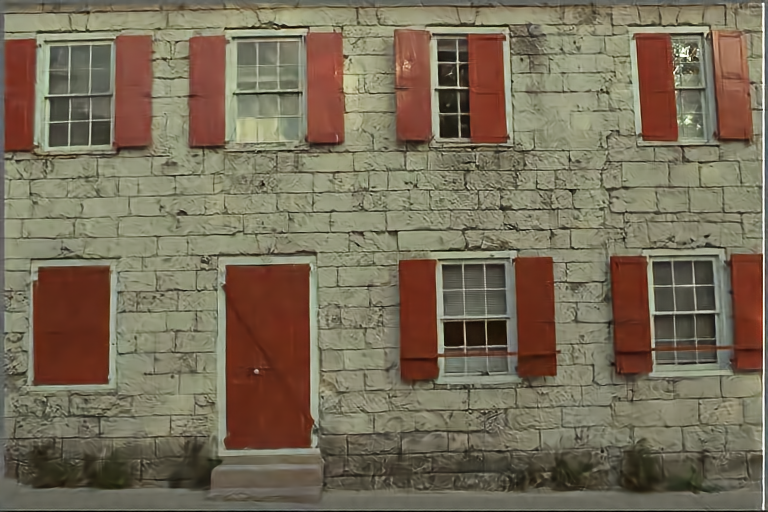} & \includegraphics[width=\ImageWidthVariable]{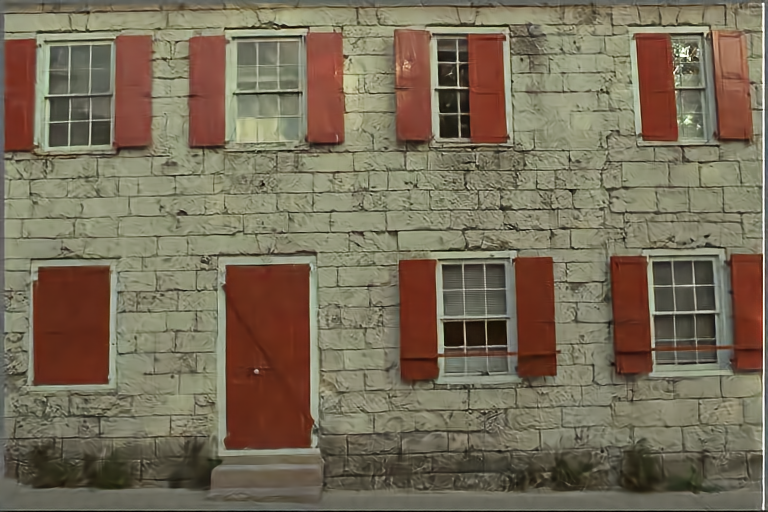} \\
    ABC & BPP=0.36,PSNR=27.39 & BPP=0.35,PSNR=27.74 & BPP=0.35,PSNR=27.77 \\
    & \includegraphics[width=\ImageWidthVariable]{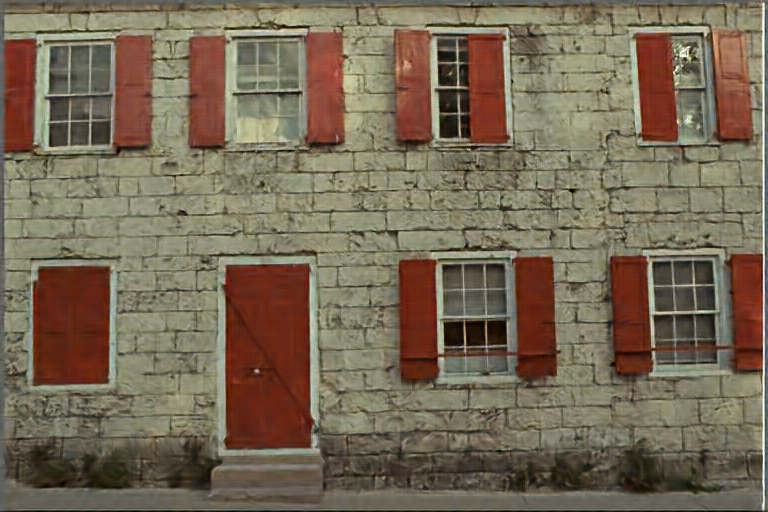} & \includegraphics[width=\ImageWidthVariable]{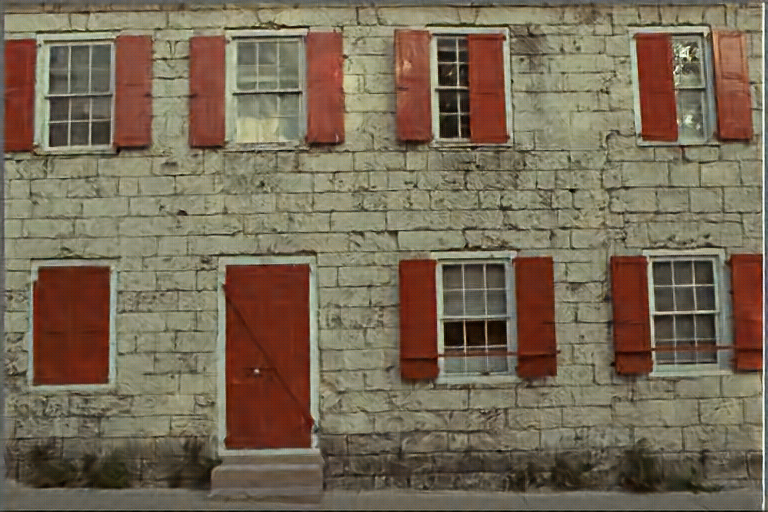} & \includegraphics[width=\ImageWidthVariable]{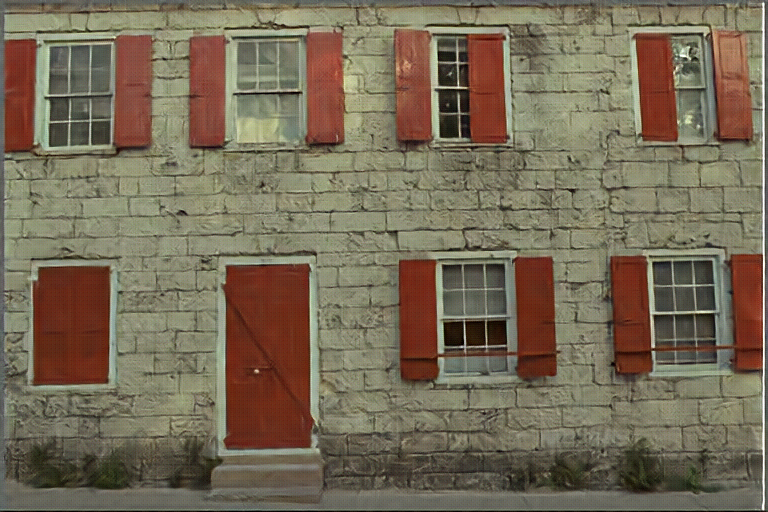} \\
    ABC-CLS & BPP=0.29,PSNR=24.83 & BPP=0.29,PSNR=24.83 & BPP=0.27,PSNR=25.06 \\
    \bottomrule
    \end{tabular}%

    }
    \label{fig:exp-sc-vis}
\end{table*}

\subsection{Comparing Lossy Image Compression Performance Under Computation Constraints}
\label{sec:exp-comp}


This section addresses the underexplored domain of computational scalability in neural image compression (NIC), focusing on lossy compression under hardware resource limitations. Existing NIC literature predominantly neglects computational adaptability; thus, we first analyze computational complexity across prominent works and select representative models aligned with predefined operational constraints. 

\begin{figure*}[t]
\centering
    \subfloat[]{
        \centering
        \includegraphics[width=.4\textwidth]{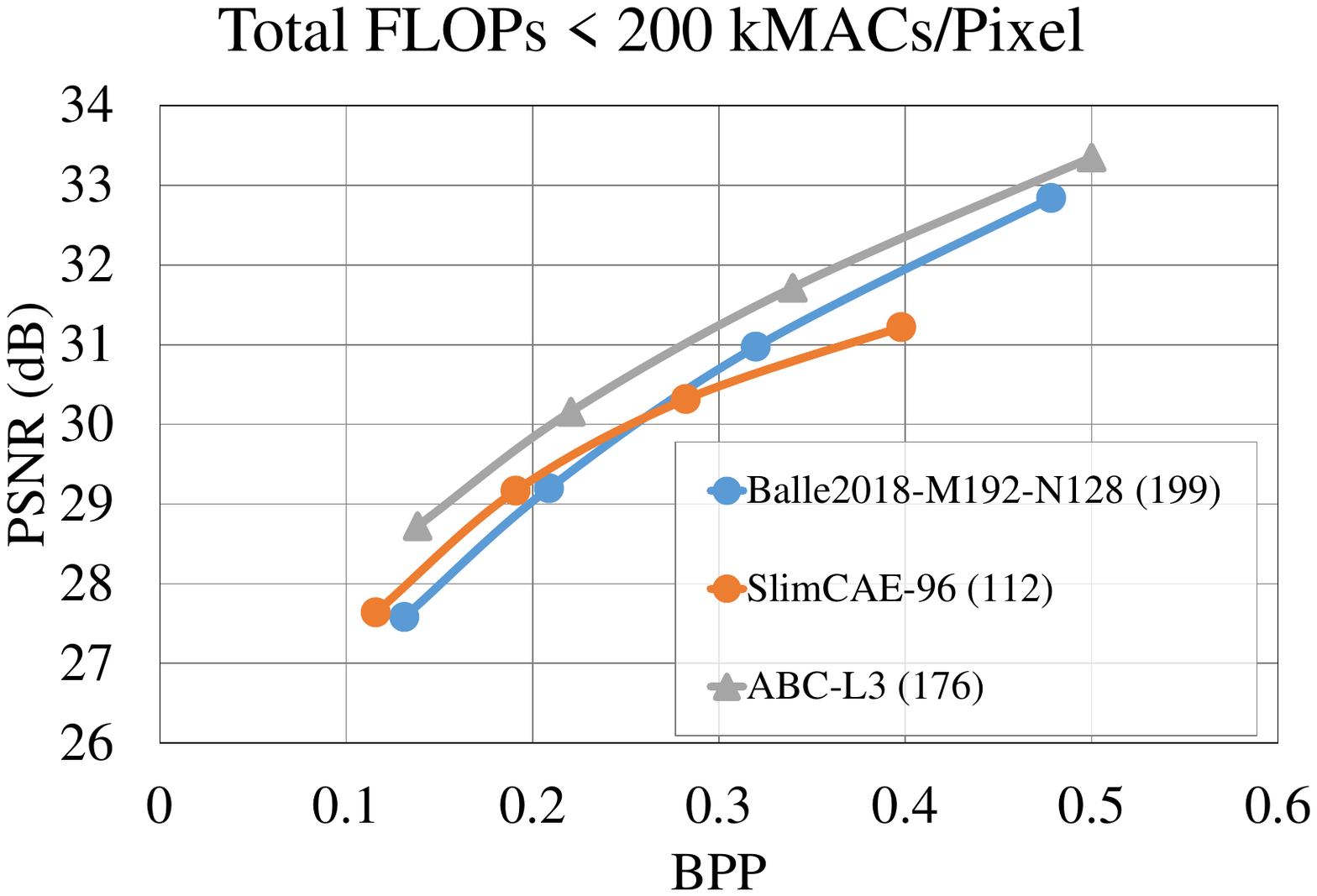}
        \label{fig:exp-comp-limflop-low}
    }
    \subfloat[]{
        \centering
        \includegraphics[width=.4\textwidth]{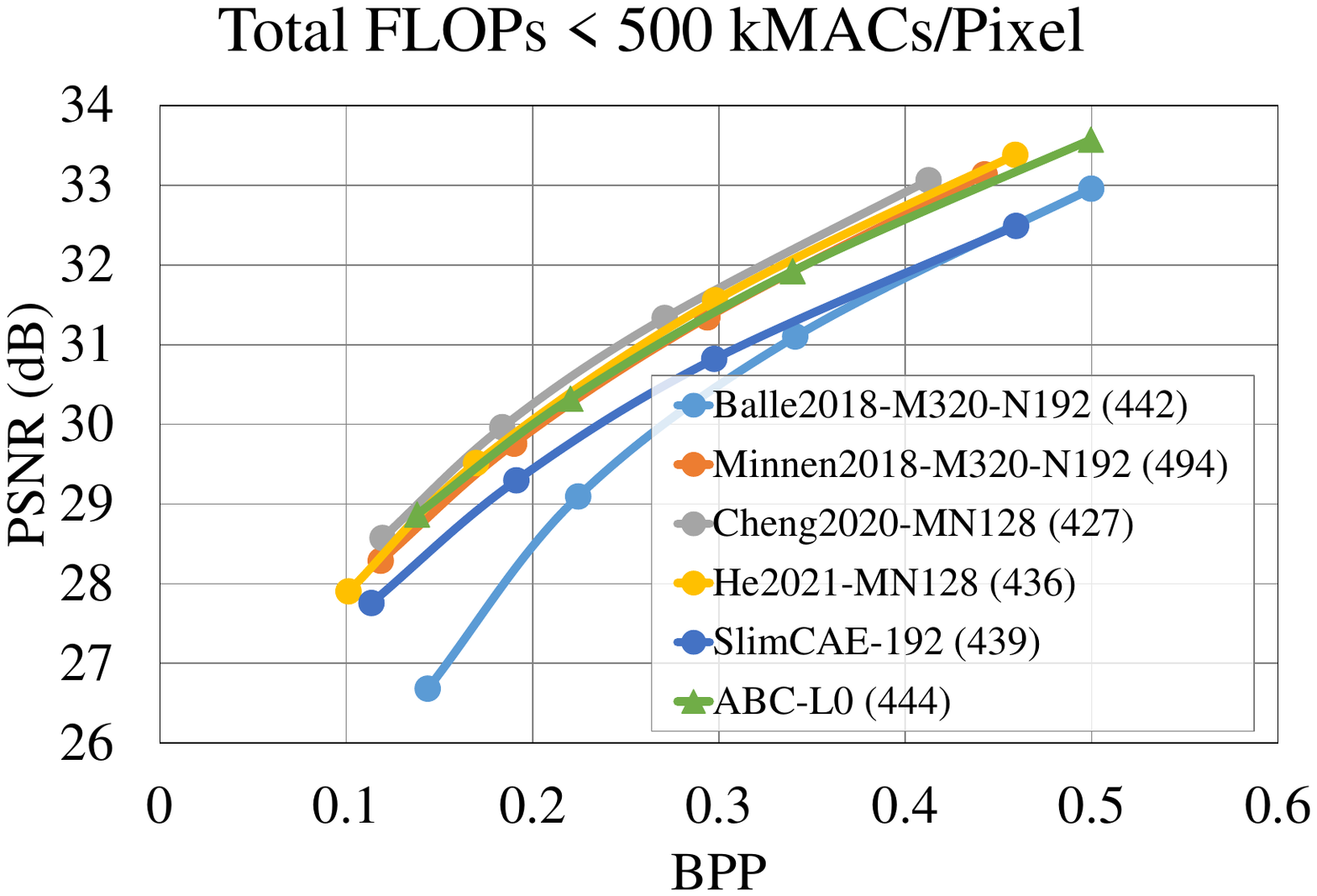}
        \label{fig:exp-comp-limflop-high}
    }
    \caption{Comparative results subject to total MACs constraints. The parenthetical value following each method in the legend denotes its MACs value in kMACs/Pixel.}
    \label{fig:exp-comp-limflop}
\end{figure*}

We evaluate our models under two distinct computational regimes: the low-complexity regime (up to 200 kMACs/pixel), which includes the baseline model Hyperprior\cite{Ball2018VariationalIC}, and the high-complexity regime (up to 500 kMACs/pixel), a range that encompasses many cutting-edge approaches. Within the low-complexity regime, we compare the low-bitrate version of Hyperprior (Balle2018-M192-N128, at 199 kMACs/pixel)\cite{Ball2018VariationalIC}, and SlimCAE with 96 channels (SlimCAE-96, at 112 kMACs/pixel)\cite{Yang2021SlimmableCA}. For the high-complexity regime, several higher complexity models serve as benchmarks: the high-bitrate version of Hyperprior (Balle2018-M320-N192, at 442 kMACs/pixel)\cite{Ball2018VariationalIC}, the high-bitrate Joint Autoregressive model (Minnen2018-M320-N192, at 494 kMACs/pixel)\cite{Ball2018VariationalIC}, the low-bitrate version of Cheng2020 (Cheng2020-MN128, at 427 kMACs/pixel)\cite{Cheng2020LearnedIC}, the low-bitrate Checkerboard Context model (He2021-MN128, at 436 kMACs/pixel)\cite{He2021CheckerboardCM}, and SlimCAE equipped with 192 channels (SlimCAE-192, at 439 kMACs/pixel)\cite{Yang2021SlimmableCA}. It is important to note that we reimplemented SlimCAE within our framework without $\lambda$-scheduling to maintain a consistent MAC across all bitrate levels. The comparative rate-distortion performance is illustrated in \autoref{fig:exp-comp-limflop}.


In the low-complexity regime, our method surpasses both Hyperprior and SlimCAE under identical MAC budgets. SlimCAE’s limited adaptability (5 discrete channel widths) restricts its operational flexibility, whereas our framework enables independent channel adjustments across four structurally identical networks, yielding $5^4 = 625$ configurable complexity levels for precise constraint adherence.
For the high-complexity regime, while outperforming SlimCAE and Hyperprior, our approach exhibits marginally inferior performance to specialized architectures like Cheng2020 and Checkerboard Context. This discrepancy likely stems from their optimized network topologies and extended training protocols, as explored in further architectural ablation studies in \autoref{sec:exp-abl-inter}.

\subsection{Comparison to Other Autoregressive Models for Intra-Node BayesNet}
\label{sec:exp-intra}

This section evaluate the effectiveness of the proposed intra-node BayesNet structure learning method, by comparing it to other hand-crafted autoregressive models. For a simple and fair comparison, we first establish a baseline using a joint autoregressive framework\cite{Minnen2018JointAA} with a null context model, trained for about 500k iterations. Subsequently, we retain all other components while training various autoregressive models to replace the baseline context model for an additional 500k iterations. This approach ensures the reconstructed image remains unchanged, allowing us to isolate and evaluate the bit-rate to reflect the effectiveness of different autoregressive models. Moreover, the decompression speed is also reported to evaluate the efficiency of
It's important to note that the neural network architecture for the context model is kept constant across all methods, with the sole variability stemming from the structure of the intra-node BayesNet.

The benchmark autoregressive models include Channel-wise (CW, 2-stage)\cite{Minnen2020ChannelWiseAE}, Checkerboard\cite{He2021CheckerboardCM} (CB, 2-stage), Multi-stage spatial 2x2\cite{Lin2023MultistageSC} (MS, 4-stage) and the combined unevenly grouped channel-wise and checkerboard \cite{He2022ELICEL} (ELIC, 10-stage in total). We also train the MaskConv context model from \cite{Minnen2018JointAA} as a baseline. Our model is trained under constraints of 2, 4, and 10 parallel stages and evaluated against counterparts with equivalent stage configurations. The findings are presented in \autoref{tab:exp-abl-ar}.

Overall, our learning-based BayesNet structures consistently demonstrate slightly better performance compared to the manually-crafted models under the same parallel stages, with similar decompression speed. 
Interestingly, in each group our learned method reaches close BPP to the hand-crafted ones in the next level of parallel stage, and in the 10-stage case, our method even approaches the BPP of non-parallel MaskConv model. 
This proves that our proposed method excels in learning effective intra-node BayesNets for autoregressive models.

\begin{table*}[t]
    \centering
    \caption{Bit-rate comparison for different autoregressive modules.}
    \resizebox{.9\linewidth}{!}{
\begin{tabular}{c|cc|ccc|cc|cc}
\toprule
Stages & \multicolumn{2}{c|}{-} & \multicolumn{3}{c|}{2} & \multicolumn{2}{c|}{4} & \multicolumn{2}{c}{10} \\
\midrule
Methods & MaskConv & Baseline & CB    & CW    & Ours  & MS    & Ours  & ELIC  & Ours \\
BPP   & 0.3150  & 0.3333  & 0.3216  & 0.3222  & \textbf{0.3207 } & 0.3210  & \textbf{0.3183 } & 0.3174  & \textbf{0.3156 } \\
Speed (MB/s) & 0.32  & 345.55  & 191.23  & 173.46  & 172.57  & 107.93  & 74.67  & 17.85  & 17.32  \\
\bottomrule
\end{tabular}%

    }
    \label{tab:exp-abl-ar}
\end{table*}

\begin{figure*}[t]

\centering
    
    \subfloat[ELIC (10stage)]{
        \centering         
        \includegraphics[width=\linewidth]{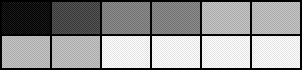}
    }

    \subfloat[Ours (10stage)]{
        \centering         
        \includegraphics[width=\linewidth]{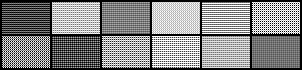}
    }

    \caption{Partite topological indices visualization for intra-node BayesNet in the 10-stage case. Each image patch indicates topological indices of nodes in a group of channels. Blacker pixels indicate lower topological indices and whiter indicate higher topological indices. E.g., in the 10-stage case, pure-black indicates index 0 which is first iterated, and pure-white indicates group 9 which is last iterated. Better viewed with zoom-in.}
    \label{fig:exp-abl-ar-topogroups}

\end{figure*}

\subsubsection{Qualitative Results}
~To elucidate the distinction between data-driven and manual architectural design, we present a comparative visualization of intra-node BayesNet topologies in \autoref{fig:exp-abl-ar-topogroups}, contrasting our learned structure with ELIC’s hand-crafted implementation under the same number of stages.
Analysis reveals that ELIC’s Bayesian network employs a checkerboard spatial partitioning scheme, dividing each channel group into two mutually exclusive partitions. In contrast, our learned architecture autonomously derives a more granular spatial dependency structure, typically partitioning channels into three or four interdependent groups. This enhanced context aggregation capacity aligns with the inherent spatial locality of natural images, where pixel correlations exhibit strong geometric regularity.
Notably, the emergent topological patterns in our learned models suggest a design principle for parallel autoregressive frameworks: systematic integration of channel-wise and spatial dependencies within unified computational graphs. This insight is also adopted by recent advances in neural video compression \cite{Li2023NeuralVC}, where interlaced channel-spatial context modeling has demonstrably improved rate-distortion efficiency, which provides empirical validation of our architectural paradigm.


\subsection{Ablation Studies for Framework Implementation}
\label{sec:exp-abl-inter}
This section include some further ablation studies for different implementation of the proposed framework described in \autoref{sec:impl}.
We first conduct a systematic evaluation of slimmable backbone networks, including architectures from Cheng2020\cite{Cheng2020LearnedIC} (configured as $M=128,N=128$ and the attention module is disabled) and ELIC\cite{He2022ELICEL}(configured as $M=192,N=192$ and the attention module is disabled), to validate our framework’s robustness across diverse neural implementations.
Subsequently, we benchmark slimmable networks at maximum channel width against non-scalable static backbone implementations. To isolate the impact of our two-stage training paradigm, we introduce a one-stage optimization variant utilizing greedy-searched complexity levels.
All experiments adhere to the experiment settings outlined for in \autoref{sec:exp-sc-mac}.

\begin{figure}[t]
    
    \centering
    \includegraphics[width=.8\linewidth]{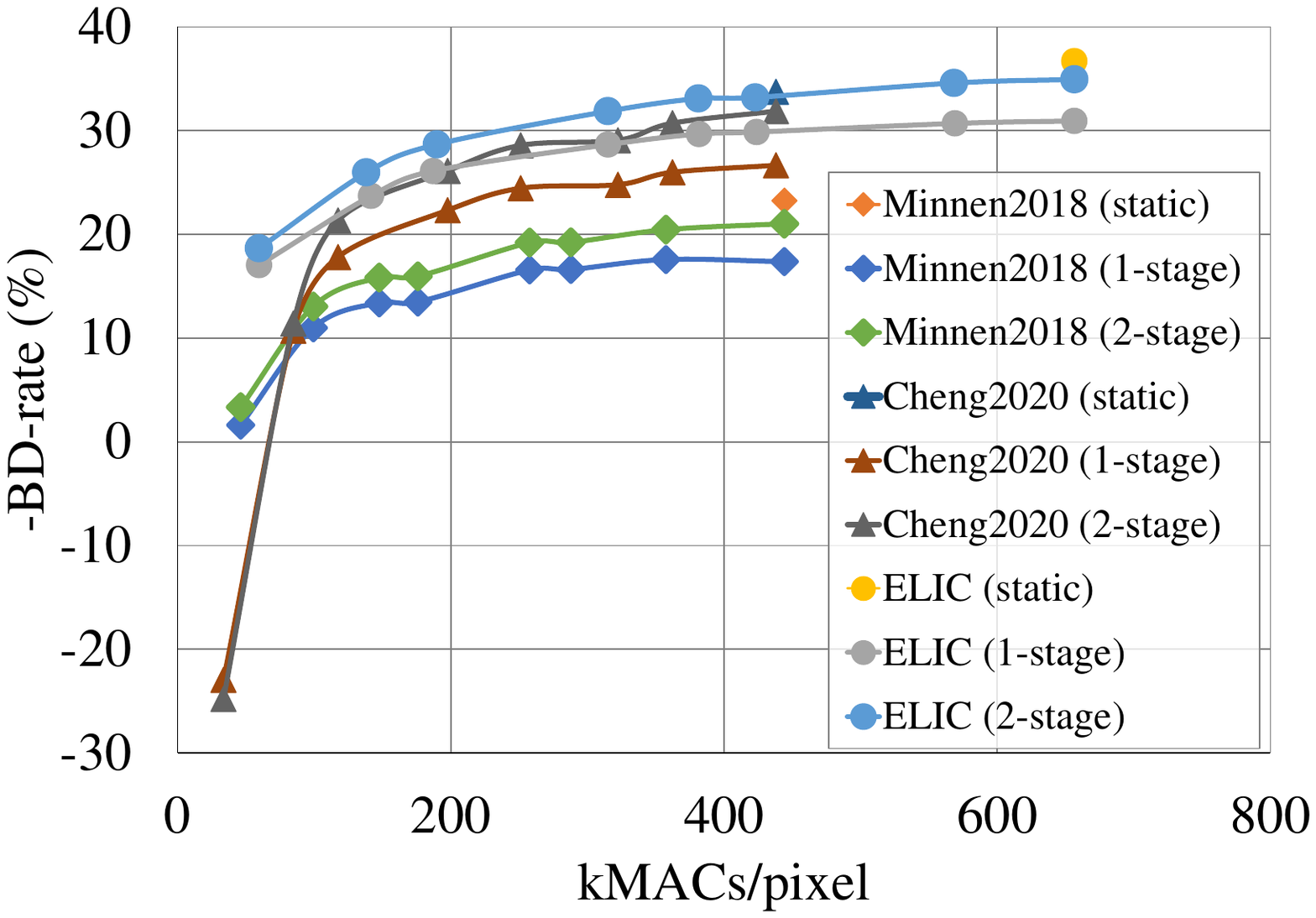}
    \caption{Ablation study results for different inter-node BayesNet implementations.}
    \label{fig:exp-abl-inter-apd}

\end{figure}

As illustrated in \autoref{fig:exp-abl-inter-apd}, our framework achieves performance comparable to static backbones under full-channel configurations, attributable to the efficacy of the two-stage optimization strategy. The one-stage variant exhibits marginally inferior results—an anticipated outcome given the inherent optimization challenges in dynamic network training.
Notably, while modern architectures (e.g., ELIC) improve compression performance by up to 15\% over baseline implementations, the core comparative trends persist universally across tested backbones. This consistency underscores our framework’s architectural agnosticism and validates its generalizability.

\section{Conclusions}
\label{sec:conc}

This paper presents a novel approach, ABC, to addressing adaptive computational scalability in NIC by focusing on adaptive BayesNet structure generation. Moreover, we decompose the challenge into two subproblems: heterogeneous bipartite graph learning for inter-node BayesNet, and homogeneous multipartite graph learning for intra-node BayesNet, proposing solutions for each. Experimental results show that our work offers improved control over computational demands relative to other scalable computation models while maintaining comparable compression performance.
\ifCLASSOPTIONcompsoc
  \section*{Acknowledgments}
\else
  \section*{Acknowledgment}
\fi

The paper is supported in part by the National Natural Science Foundation of China (No. 62325109, U21B2013).

\ifCLASSOPTIONcaptionsoff
  \newpage
\fi



\bibliographystyle{IEEEtran}
\balance
\bibliography{IEEEabrv,ref}
%




%

\begin{IEEEbiographynophoto}{Yufeng Zhang}
 received the B.E. and M.E. degrees in
 electronic engineering from Shanghai Jiao Tong University in 2018 and 2021, respectively, and is currently pursuing the PhD degree in Department of Electronic Engineering, Shanghai Jiao Tong University. His current research directions include multimedia data compression, probabilistic models, and computer vision.
\end{IEEEbiographynophoto}

\begin{IEEEbiographynophoto}{Wenrui Dai}
received B.S., M.S., and Ph.D. degree in Electronic Engineering from Shanghai Jiao Tong University, Shanghai, China in 2005, 2008, and 2014. He is currently an associate professor at the Department of Computer Science and Engineering, Shanghai Jiao Tong University (SJTU). Before joining in SJTU, he was with the faculty of the University of Texas Health Science Center at Houston from 2018 to 2019. He was a postdoctoral scholar with the Department of Biomedical Informatics, University of California, San Diego from 2015 to 2018 and a postdoctoral scholar with the Department of Computer Science and Engineering, SJTU from 2014 to 2015. His research interests include learning-based image/video coding, image/signal processing and predictive modeling.
\end{IEEEbiographynophoto}

\begin{IEEEbiographynophoto}{Hang Yu}
received the PhD degree in electrical and electronics engineering (EEE) from Nanyang Technological University (NTU), Singapore, in 2015. He is currently a senior algorithm expert with Ant Group, China, since December 2020. Before joining Ant Group, he was a senior research fellow with NTU, Singapore. His research interests include time series modeling, graph machine learning, natural language processing, and generative models.
\end{IEEEbiographynophoto}

\begin{IEEEbiographynophoto}{Shizhan Liu}
Shizhan Liu received the B.E. and M.E. degrees in electronic engineering from Shanghai Jiao Tong University in 2021 and 2024, respectively. He is currently employed at Ant Group, where his research interests focus on multimedia data compression, diffusion models, and multimodal models.
\end{IEEEbiographynophoto}


\begin{IEEEbiographynophoto}{Junhui Hou}
 received the BE and ME degrees in
 communication engineering from Huaqiao University, Xiamen, China, in 2017 and 2020, respectively.
 He is currently working toward the PhD degree in
 computer science with the City University of Hong
 Kong. His research interests include hyperspectral
 image processing and deep learning.
\end{IEEEbiographynophoto}

\begin{IEEEbiographynophoto}{Jianguo Li}
 received the Ph.D. degree from Tsinghua
 University, Hangzhou, China, in July 2006. Since June
 2020, he has been a Senior Staff Engineer/tech
Director with Ant Group. After the Ph.D. degree,
 he was a Research Scientist with Intel Labs for near
 14 years. He has authored or coauthored about 50
 peer-reviewed papers in top-tier conferences or jour
nals, and hold more than 50 U.S. issued patents.
 His research interests include machine learning, deep
 learning, and their applications. He also led team to
 win or perform top on various well-known vision
 challenges, including TRECVID, Middlebury MVS, UMass FDDB, PASCAL
 VOC, MSR-VTT, VQA2.0, and NIPS 2017 adversarial vision challenge.
\end{IEEEbiographynophoto}

\begin{IEEEbiographynophoto}{Weiyao Lin}
 received the B.E. and M.E. degrees in
 electrical engineering from Shanghai Jiao Tong Uni
versity, Shanghai, China, in 2003 and 2005, respec
tively, and the Ph.D. degree in electrical and com
puter engineering from the University of Washing
ton, Seattle, WA, USA, in 2010. He is currently a
 Professor with the Department of Electronic Engi
neering, Shanghai Jiao Tong University. His current
 research interests include image/video processing,
 video surveillance, and computer vision.
\end{IEEEbiographynophoto}




    
\clearpage

\appendices
    
\section{Notation List}
\label{sec:apd-notation}
See \autoref{tab:notation}.

\begin{table*}[]
    \centering
    \caption{Symbols used in this paper.}
    \begin{tabularx}{\textwidth}{c|X}
    $\rvx$     &  Data/Observation node. \\
    $\rvy$     &  Latent node. \\
    $\rvz$     &  Hyper-latent node. \\
    $\hat{\rvy}$     &  Samples of $\rvy$. Note that all symbols with a hat stands for samples. \\
    $\rmX$     &  Data/Observation node set. \\
    $\rmL$     &  Latent node set. \\
    $\rmG$     &  BayesNet structure node set. \\
    $\mPhi$     &  External controller node for generating BayesNet. \\
    $f(\rmG), f(\rmG, \rmL)$     &  Optimization target function. \\
    $\Ls_{R}, \Ls_{D}, \Ls_{C}$     &  Loss function for rate, distortion and complexity. \\
    $\lambda_{D}, \lambda_{C}$     &  Hyperparameters to adjust trade-off between rate, distortion and complexity. \\
    $\rmG_{inter}$     & Inter-node BayesNet structure node set. \\
    $\rmG_{intra}$     & Intra-node BayesNet structure node set. \\
    $\rmG^{\rvy, \rvz}_{inter}$     & Inter-node BayesNet structure node for $p(\rvy | \rvz)$ (or $q(\rvy | \rvz)$). \\
    $\pi$     &  Logit parameters for Bernoulli/Categorical distribution. \\
    $\ermT_{c,h,w}$     &  Multipartite graph topological index node on channel dimension $c$, height dimension $h$ and width dimension $w$. \\
    $\rmG^{i}_{intra}$     & The i-th Monte-Carlo sample of $\rmG_{intra}$  \\
    $g(\rmN)$     & Unconditional generative model (for intra-node BayesNet) \\
    $\etW_{c,h,w}$     & Dynamic convolution kernel weight to convolute with the input on channel dimension $c$, height dimension $h$ and width dimension $w$.\\
    $g_{a}, g_{s}, h_{a}, h_{s}$     & Analysis, synthesis, hyper-analysis, hyper-synthesis models, respectively. Modules defined in \cite{Ball2018VariationalIC}. \\
    
    \end{tabularx}
    \label{tab:notation}
\end{table*}



\section{Further Implementation Details}
\label{sec:apd-impl}

\subsection{Intra-node BayesNet Implementation}
\label{sec:apd-impl-intra}

\subsubsection{Computational Complexity of Multipartite-based Dynamic Masked Convolution}~
Building on the technique outlined in \autoref{sec:method-intra}, \autoref{eq:impl-ar-dynconv} and \autoref{eq:impl-ar-maskconv} provide an instantiation of intra-node BayesNet via multipartite-based dynamic masked convolution.
Here we further provide a PyTorch\footnote{\url{https://pytorch.org/}}-style pseudo-code implementation in \autoref{alg:dynconv} to clarify the implementation details.
Measuring with MAC, the complexity of the multipartite-based dynamic masked convolution could be calculated by counting the MACs in convolution operations by:
\begin{equation}
\label{eq:dynconv-mac}
\begin{aligned}
    \MAC_{\dynconv} &= C \times H_{out} \times W_{out} \\
    &\times ((C_{out} // C) \times (C_{in} // C) \times C \times K \times K) \\
    &= H_{out} \times W_{out} \times C_{out} \times C_{in} \times K \times K
\end{aligned}
\end{equation}
We could see that the proposed operator has equivalent parameter count and computational complexity to that of standard 2D convolution operators. In practice, however, it usually runs slower on GPUs compared to traditional 2D convolutions, which is partly attributable to our current implementation not being fully GPU-optimized.

\begin{algorithm}
\caption{Pseudocode for Multipartite-based Dynamic Masked Convolution.}
\label{alg:dynconv}

\begin{algorithmic}
\Require  Kernel size $K$, Output channels $C_{out}$, Input channels $C_{in}$, Dynamic channel groups (or channels of topological indices) $C$, Input tensor $\hat{\rvy} \in \R^{C_{in} \times H_{in} \times W_{in}}$, Topological indices $ \hat{\mT} \in \R^{C \times H_{in} \times W_{in}} $, Kernel weight $\tW \in \R^{C_{out} \times C_{in} \times K \times K}$, Kernel bias $\tB \in \R^{C_{out}}$,
\Ensure Output tensor $\hat{\rvy}_{out} \in \R^{C_{out} \times H_{out} \times W_{out}}$

\State Reshape $\hat{\rvy} \in \R^{(C_{in} // C) \times C \times H_{in} \times W_{in}}$
\State Reshape $\tW \in \R^{C \times (C_{out} // C) \times (C_{in} // C) \times C \times K \times K}$
\State Reshape $\tB \in \R^{C \times (C_{out} // C)}$
\State Initialize $\hat{\rvy}_{out} \in \R^{(C_{out} // C) \times C \times H_{out} \times W_{out}} $

\For{$c,h,w$ in range($C,H,W$)} 

\State Unfold2D $\hat{\rvy}_{knl} \in \R^{(C_{in} // C) \times C \times K \times K} $

\State Unfold2D $\hat{\emT}_{knl} \in \R^{1 \times C \times K \times K}$

\State Get mask $\hat{\rvy}_{mask} \gets Binarize(\hat{\emT}_{knl} > \hat{\emT}[c,h,w])$
\Comment{$\hat{\rvy}_{mask} \in \R^{1 \times C \times K \times K}$}

\State Apply mask $\hat{\rvy}_{input} \gets (\hat{\rvy}_{mask} \cdot \hat{\rvy}_{knl}) $
\Comment{$\hat{\rvy}_{input} \in \R^{(C_{in} // C) \times C \times K \times K}$}

\State Convolution $\hat{\rvy}_{conv} \gets \sum^{(C_{in} // C),C,K,K} (\hat{\rvy}_{input} \cdot \etW[c]) + \tB[c] $ 
\Comment{$\hat{\rvy}_{conv} \in \R^{(C_{out} // C)}$}

\State Output $\hat{\rvy}_{out}[:,c,h,w] \gets \hat{\rvy}_{conv}$

\EndFor

\State Reshape $\hat{\rvy}_{out} \in \R^{C_{out} \times H \times W}$

\end{algorithmic}

\end{algorithm}

\subsubsection{Merging Predictions from Intra-node and Inter-node BayesNet}~
In the original framework, the complete prior for predicting $\hat{\rvy}$ given $\hat{\rvz}$, denoted as $p(\hat{\rvy} | \hat{\rvz})$, fuses the output from $h_{s}(\hat{\rvz})$ and the context model $\AR(\hat{\rvy})$. This is achieved using a 3-layer MLP, consisting of $1 \times 1$ convolutional layers and LeakyReLU activation, as follows:
\begin{equation}
    p(\hat{\rvy} | \hat{\rvz}) \propto \MLP(\concat(h_{s}(\hat{\rvz}), \AR(\hat{\rvy}))).
\end{equation}
In our approach to formulating the complete prior, $p(\hat{\rvy} | \hat{\rvz}, \rmG^{\hat{\rvy},\hat{\rvz}}_{inter}, \rmG^{\hat{\rvy}}_{intra})$, a similar 3-layer network is applied for merging. However, the nature of our autoregressive process across the channel dimension precludes the use of standard $1 \times 1$ convolutions. Instead, we employ the dynamic masked convolution with $1 \times 1$ kernels, accommodating autoregressive operations over three dimensions. Importantly, the output from $h_{s}(\hat{\rvz}, \rmG^{\hat{\rvy},\hat{\rvz}}_{inter})$ is treated as the initial partite, preceding the context model’s output $\AR(\hat{\rvy}, \rmG^{\hat{\rvy}}_{intra})$ because $\hat{\rvz}$ has topological precedence over $\hat{\rvy}$ in the BayesNet structure. This is represented as:
\begin{equation}
    \begin{aligned}
    &p(\hat{\rvy} | \hat{\rvz}, \rmG^{\hat{\rvy},\hat{\rvz}}_{inter}, \rmG^{\hat{\rvy}}_{intra}) \propto \MLP_{dyn}(
    \concat(h_{s}(\hat{\rvz}, \rmG^{\hat{\rvy},\hat{\rvz}}_{inter}),  \\
    &\AR(\hat{\rvy}, \rmG^{\hat{\rvy}}_{intra})), \concat(\hat{\emT}^{\hat{\rvy}}, \hat{\emT}^{\hat{\rvz}})), \\
    &\hat{\emT}^{\hat{\rvy}} \in \{0,1,\ldots,S_{intra}\}, \quad \hat{\emT}^{\hat{\rvz}} \in \mathbf{-1}.
    \end{aligned}
\end{equation}
Our merging neural network, when compared to \cite{Minnen2018JointAA}, maintains the same MACs and parameters for equivalent channel configurations, as per \autoref{eq:dynconv-mac}. Additionally, when contrasted with channel-wise autoregressive models\cite{Minnen2020ChannelWiseAE}, which employ several merging networks for channel groups, our method is also more MAC and parameter efficient, as the $h_{s}$ prediction is processed solely once.

\subsubsection{Learning the Intra-node BayesNet Generator}~
Earlier in \autoref{sec:method-intra}, we outlined the use of a generative model $g(\rmN)$ in \autoref{eq:ga-dist-approx} to model the latent correlations among nodes in the intra-node BayesNet. This model is a straightforward 2-layer network comprising Linear and ReLU layers. Nonetheless, we encountered optimization challenges leading to local optima that resulted in ineffective BayesNets. This is often due to the most effective intra-node BayesNets having diverse topological indices for adjacent nodes, which increases the divergence of their corresponding distributions and causes the Monte-Carlo sampler to converge prematurely.

To address this, we employed two strategies in optimizing the generative model. Firstly, we designed $g(\rmN)$ to output logits for a $C \times 2 \times 2 \times S_{intra}$ tensor representing the logits of $\prod^{C,H=2,W=2}{c,h,w} p(\ermT{c,h,w})$. From this distribution, we sample $C \times 2 \times 2$ topological indices and then expand them over the spatial dimensions to form a $C \times H \times W$ tensor of $\hat{\emT}_{c,h,w}$. This approach effectively makes every node within the $2 \times 2$ tensor adjacent, thereby reducing the likelihood of optimization collapse.

Secondly, to circumvent the collapsing issue and locate a more favorable starting point for optimization, we initially train a context model with random characteristics akin to the approach in \cite{He2021CheckerboardCM}. Instead of utilizing a random mask, we employ random topological indices for $\hat{\emT}{c,h,w}$. This preliminary training phase lasts about 100,000 iterations without including $\Ls{\rmG_{intra}}$ in the loss function. Subsequently, we initiate the Monte-Carlo optimization by incorporating $\Ls_{\rmG_{intra}}$, guiding the model towards improved local optima.

\subsection{Training and Evaluation}
\label{sec:apd-impl-trainval}

\subsubsection{Optimization}~
Recall that we use the loss function \autoref{eq:impl-loss}:
\begin{equation}
\label{eq:impl-loss-recall}
    \Ls = \Ls_{R} + \lambda_{D}\Ls_{D} + \lambda_{C} \Ls_{C} + \Ls_{\rmG_{intra}}
\end{equation}
Specifically, the rate loss, $\Ls_{R}$, integrates all latent variables within $\rmL$ as: 
\begin{equation}
    \begin{aligned}
    &\Ls_{R} = \\
    &\E_{\hat{\rvy} \sim q(\rvy | \rvx, \rmG^{\rvy,\rvx}_{inter}), \hat{\rvz} \sim q(\rvz | \rvy, \rmG^{\rvz,\rvy}_{inter})} \log_{2} p(\hat{\rvy} | \hat{\rvz}, \rmG^{\hat{\rvy},\hat{\rvz}}_{inter}, \rmG^{\hat{\rvy}}_{intra}) p(\hat{\rvz}).
    \end{aligned}
\end{equation}
Distortion loss using MSE is given by: 
\begin{equation}
    \Ls_{D} = || \hat{\rvx} - \rvx ||_{2} , \quad \hat{\rvx} = \argmax_{\hat{\rvx}} p(\hat{\rvx} | \hat{\rvy}, \rmG^{\hat{\rvx},\hat{\rvy}}_{inter}),
\end{equation}
The complexity loss then encompasses the complexity of all BayesNets: 
\begin{equation}
    \Ls_{C} = C(\rmG^{\rvy,\rvx}_{inter}) + C(\rmG^{\rvz,\rvy}_{inter}) + C(\rmG^{\hat{\rvy},\hat{\rvz}}_{inter}) + C(\rmG^{\hat{\rvx},\hat{\rvy}}_{inter}) + C(\rmG^{\hat{\rvy}}_{intra}).
\end{equation}
where $\Ls_{C}(\rmG_{inter})$ can be computed as in \autoref{eq:bn-latent-hetero-complexity}. 
Given that masked convolution is implemented as per \autoref{alg:dynconv}, $\Ls_{C}(\rmG^{\hat{\rvy}}_{intra})$ is constant and thus omitted from the loss function. 
Finally, with $\rmG_{intra}$ applied only to $\hat{\rvy}$ in this instance, the VIMCO loss for optimizing $\rmG_{intra}$ is:
\begin{equation}
    \Ls_{\rmG_{intra}} = \E_{q(\rmG^{\hat{\rvy},1:K}_{intra})} \log \frac{1}{K} \sum^{K}_{i=1} \log p(\hat{\rvy} | \hat{\rvz}, \rmG^{\hat{\rvy},\hat{\rvz}}_{inter}, \rmG^{\hat{\rvy}, i}_{intra})
\end{equation}

\subsubsection{Training Schedule}~
The training stage spans approximately 2M iterations (1.6M in the first stage and 0.4M in the second stage), employing the Adam optimizer with a learning rate of 0.0001. 
We follow CompressAI by employing four different distortion weights, $\lambda_{D} = [0.0018,0.0035,0.0067,0.0130]$ for PSNR and $[2.40,4.58,8.73,16.64]$ for MS-SSIM, allowing us to craft a suite of models that cater to a spectrum of bit-rate preferences. Note that the MS-SSIM models are fine-tuned from pretrained PSNR models.


\subsubsection{Complexity Metrics}~
We utilize ptflops\footnote{\url{https://github.com/sovrasov/flops-counter.pytorch}} to tally the MAC figures. For decompression speed, we measure by calculating the ratio of the total size of the decompressed image data (in MegaBytes) to the aggregate time taken (in seconds) to decompress the entire image dataset. It's important to note that our image-loading routine is grounded in torchvision\footnote{\url{https://pytorch.org/vision/stable/index.html}}, which, by standard practice, loads images into 32-bit floating point tensors. Consequently, the volume of bytes in the decompressed images is quadrupled, mirroring the dimension count. As a result, the decompression speeds we report are effectively quadruple those cited in typical benchmarks.


\section{Further Experimental Results}
\label{sec:apd-exp}

\subsection{Ablation Studies for Intra-node BayesNet Implementation}
In this section, we delve into additional ablation studies concerning the intra-node BayesNets, focusing on key hyperparameters related to homogeneous multipartite graph learning, such as the number of channel groups $C$ and the kernel size used in dynamic masked convolutions. As in \autoref{sec:exp-intra}, we maintain a frozen backbone while training the learnable BayesNets and evaluate each variation based on BPP and decompression speed. Moreover, we provide visual representations of the partite topological indices. These results are documented in \autoref{tab:exp-abl-intra-apd}.

\begin{table}[t]
    \centering
    \newcommand{\customwidth}{.35\linewidth}
    \caption{Ablation study results for different intra-node BayesNet implementations.}
\begin{tabular}{c|cc|ccc}
\toprule
Stages & $C$   & $K$   & BPP   & Speed & Vis \\
\midrule
\multirow{4}[2]{*}{2} & 2     & 5     & \textbf{0.3105} & 170.30  &  \includegraphics[width=\customwidth]{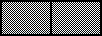} \\
      & 4     & 5     & 0.312 & 142.02  &  \includegraphics[width=\customwidth]{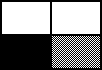} \\
      & 2     & 3     & 0.3135 & 193.41  &  \includegraphics[width=\customwidth]{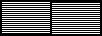} \\
      & 2     & 7     & 0.312 & 153.93  &  \includegraphics[width=\customwidth]{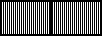} \\
\midrule
\multirow{3}[2]{*}{4} & 2     & 5     & \textbf{0.3075} & 96.72  &  \includegraphics[width=\customwidth]{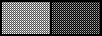} \\
      & 4     & 5     & \textbf{0.3075} & 80.96  &  \includegraphics[width=\customwidth]{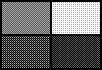} \\
      & 6     & 5     & 0.3081 & 69.21  &  \includegraphics[width=\customwidth]{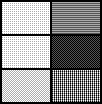} \\
\bottomrule
\end{tabular}%

    \label{tab:exp-abl-intra-apd}

\end{table}


The data indicates that for a 2-stage structure, $C=2$ yields better results than $C=4$. In a 4-stage configuration, both $C=2$ and $C=4$ deliver comparable outcomes, while $C=6$ underperforms slightly. The choice of $C$ appears to have a modest effect on performance; however, aligning $C$ with the number of parallel stages may be advantageous.Examining the visualized graphs, the arrangement for $C=2$ resembles a checkerboard context model with a distinct channel-wise distribution of the black and white patterns—a formation we could term an "interlaced checkerboard." Meanwhile, the graph for $C=4$ presents a hybrid of checkerboard and channel-wise autoregressive patterns, which is less effective than its interlaced counterpart.


When it comes to the kernel size, neither the $3 \times 3$ nor the $7 \times 7$ options surpass the performance of the widely implemented $5 \times 5$ kernels, a finding consistent with observations from the MaskConv context model\cite{Minnen2018JointAA}. Although without good performance, the learned topological indices also showcases an interlaced pattern, which indicate its effectiveness. 


In summary, the optimal selection for the number of channel groups, $C$, generally corresponds to the number of processing stages, while the kernel size, $K$, should typically be set to 5. Most of the learned intra-node BayesNets exhibit an interlaced pattern, irrespective of the hyperparameters. This pattern is likely to capture more comprehensive context information by integrating cues from both the channel and spatial dimensions. Interestingly, the interlaced approach resonates with the concepts presented in ELIC\cite{He2022ELICEL}, which amalgamates channelwise autoregressive models with a checkerboard context model. This convergence of strategies serves to underscore the efficacy of our proposed learning-based autoregressive model and offers valuable direction for the development of context models in the field of NIC.






\end{document}